\documentclass{pasj00}

\Received{$\langle$reception date$\rangle$}
\Accepted{2016 January 11}
\Published{$\langle$publication date$\rangle$}
\SetRunningHead{Astronomical Society of Japan}{Usage of \texttt{pasj00.cls}}

\usepackage{times}
\usepackage{url}

\usepackage{lineno}
\modulolinenumbers[1]

\begin{document}

\title{MAXI observations of long-term variations of Cygnus X-1 in the low/hard and the high/soft states}
\author{Juri Sugimoto$^{1,2}$, Tatehiro Mihara$^1$, Shunji Kitamoto$^{2,5}$, Masaru Matsuoka$^1$, Mutsumi Sugizaki$^1$,\\
Hitoshi Negoro$^3$, Satoshi Nakahira$^4$ and Kazuo Makishima$^1$
\thanks{Last update: Jan 9, 2016}}
\affil{%
$^1$  MAXI team, Institute of Physical and Chemical Research (RIKEN), 2-1 Hirosawa, Wako, Saitama 351-0198, Japan\\
$^2$  Department of Physics, Rikkyo University, 3-34-1 Nishi-Ikebukuro, Toshima, Tokyo 171-8501, Japan\\
$^3$  Department of Physics, Nihon University, 1-8-14, Kanda-Surugadai, Chiyoda-ku, Tokyo 101-8308\\
$^4$  ISS Science Project Office, Institute of Space and Astronautical Science (ISAS),\\
 Japan Aerospace Exploration Agency (JAXA), 2-1-1 Sengen, Tsukuba, Ibaraki 305-8505, Japan\\
$^5$  Research Center for Measurement in Advanced Science,\\
 Rikkyo University, 3-34-1 Nishi-Ikebukuro, Toshima, Tokyo 171-8501, Japan
}
\email{sugimoto@crab.riken.jp}

\maketitle

\begin{abstract}

Long-term X-ray variability of the black hole binary, Cygnus X-1, 
was studied with five years of MAXI data from 2009 to 2014, which include 
substantial periods of the high/soft state, as well as the low/hard state. 
In each state, Normalized Power Spectrum densities (NPSDs) were calculated in three energy bands of 2–-4 keV, 4–-10 keV and 10-–20 keV. 
The NPSDs in a frequency from $10^{-7}$ Hz to $10^{-4}$ Hz are all approximated by a power-law function with an index $-1.35\sim -1.29$.
The fractional RMS variation $\eta$,
calculated in the above frequency range,
was found to show the following three properties;
(1) $\eta$ slightly decreases with energy in the low/hard state;
(2) $\eta$  increases towards higher energies in the high/soft state;
and (3) in the 10--20 keV band, $\eta$ is 3 times higher
in the high/soft state than in the low/hard state.
These properties were confirmed through studies of
intensity-correlated changes of the MAXI spectra.
Of these three findings,
the first one is consistent with that seen in the short-term variability during the LHS.
The latter two can be understood as a result of high variability
of the hard-tail component seen in the high/soft state with the above very low frequency range,
although the origin of the variability remains inconclusive.

\end{abstract}

\section{Introduction}
Black-hole (BH) X-ray binaries emit X-rays through the mass accretion from their companion stars.
The matter accretes onto the BH forming an accretion disk, 
which is considered to work as an efficient energy-release engine.  
The properties of accretion disks have been studied by both theoretical and observational approaches.  
From the theoretical studies, a disk is considered to evolve from the RIAF (Radiative Inefficient Accretion Flow ; \cite{Ichimaru77, Narayan94}) state, 
through the standard-disk state \citep{shakura}, up to the slim-disk state
\citep{abramowicz88}, as the accretion rate increases.  
Observationally, galactic BH binaries show two spectral states, 
the low/hard state (hereafter LHS) which is dominated  by a power-law spectrum, possibly corresponding to the RIAF,  
and the high/soft state (hereafter HSS) which is dominated by an optically-thick thermal emission from an accretion disk, i.e. the standard-disk \citep{tanaka,remillard2006,done}.
In the HSS, a hard tail component is seen in the energy spectrum, which often extends from 10 keV to several MeV by a power-law \citep{gierlinski}.
In the LHS, there is also hard tail extending $\sim 100$ keV with a clear cut-off,
whose origin is considered that thermal Comptonization of disk photons by hot plasma.
The origin of hard tail in the HSS is different from that in the LHS, and is still unknown.

Time variations of the X-ray intensity are another important aspect of BH X-ray binaries.
Especially Cygnus X-1 (hereafter Cyg X-1), the leading galactic BH binary with an orbital period of 5.6 d,
has provided rich information on the short-term variability in both states (e.g. \cite{chura, grinberg}).
The power spectrum density (PSD) in the LHS has a power-law shape with an index of about $-1$ in the frequency range from 10$^{-4}$ to 10$^{-2}$ Hz.
Between 10$^{-2}$ Hz and 10$^{-1}$ Hz, the PSD is constant as a white noise. 
Above 10$^{-1}$ Hz, the index returns to about $-1$.
There is a break again at around 1 Hz, above which the index becomes steeper (e.g., \cite{belloni, nowak2000, negoro, pot}).
These time scales are thought to reflect the dynamics of accretion flows, and the process of 
energy dissipation or particle acceleration in a vicinity of the BH.
The break at 1 Hz in the LHS can be interpreted as a decay time scale of individual shots, which appear in a light curve \citep{negoro}.
In the HSS, the PSD is approximated by a power-law also with an index of  $-1$,
but this form continues, without flattening, from $10^{-4}$ Hz up to 20 Hz, beyond which it steepens.
The PSD extending with a constant index $-1$ suggests that the accretion flows in the HSS have no characteristic time scales over the $10^{-4}$ -- 20 Hz range.
\citet{chura} explained the overall shape of the PSD in the LHS and the HSS of Cyg X-1 with 
a phenomenological model that the optically-thick disk is sandwiched by optically-thin accretion flows extending up to a large distance from the BH. 
In the LHS the optically-thick disk is truncated at some distances from the BH, 
turning into optically-thin and geometrically thick accretion flows \citep{makishima}.

On frequencies lower than $\sim 10^{-4}$ Hz (time scales of 0.1 d), 
the variability of Cyg X-1 has been studied mainly by applying PSD analyses to the data obtained with RXTE observations. 
\citet{reig} calculated PSDs of Cyg X-1 using more than eleven years of data obtained with the RXTE/ASM.
They showed that PSDs in the $10^{-7}$ to $10^{-5}$ Hz range can be expressed by a power-law with an index of $-1$, 
and that the PSDs depend on the energy (over 1.3 -- 12.2 keV) 
by a factor of 3 $\sim$ 5.
The long-term variability of hard X-rays from Cyg X-1 was studied by \citet{Vikhlinin94}, down to 4$\times$10$^{-4}$ Hz, using  the GRANAT/SIGMA data.
They detected a very low frequency QPO (Quasi-Periodic Oscillation) around  0.04--0.07 Hz.
However, on frequencies of $\ltsim 10^{-4}$ Hz, our understanding of Cyg X-1 variability has remained much poorer, 
especially in the HSS, 
than the rich information accumulated on the short-term variability.

The Monitor of All-sky X-ray Image (MAXI) obtained a long-term light curve of Cyg X-1 for more than five years.
The object had been in the LHS until 2010 June, 
and after that it stayed mainly in the HSS \citep{Negoro10}. 
By analyzing PSDs and energy spectra from this unique data set, we investigated the characteristics of long-term variations of Cyg X-1 in both states.

\section{Observation}
\label{sec:obs}
The data of Cyg X-1 to be analyzed here were obtained with MAXI (\cite{matsuoka}), 
which is attached to the International Space Station.
As the International Space Station orbits the earth in every 92 min, MAXI scans over nearly the entire sky with two kinds of X-ray cameras: 
the Gas Slit Camera (GSC: \cite{mihara}) covering the energy band of 2--20 keV,
and the Solid-state Slit Camera (SSC: \cite{tomida}) covering 0.7--7 keV.

From the MAXI home page \footnote{\url{http://maxi.riken.jp/top/index.php?cid=1&jname=J1958+352}}, we downloaded one-day bin and 90-minute bin archival data of Cyg X-1.
These archival data were selected on the condition that X-ray incident angle to a camera is smaller than 38 $^\circ$ and scan times of the source and the background are both longer than 15 s.
Energy spectra of the GSC and SSC were processed by the MAXI on-demand data web page\footnote{\url{http://maxi.riken.jp/mxondem/}} (\cite{nakahira}).
We selected $2^{\circ}.0$ for a source region and $3^{\circ}.0$ for a background region.

\begin{figure*}
  \begin{center}
   \includegraphics[width=16cm]{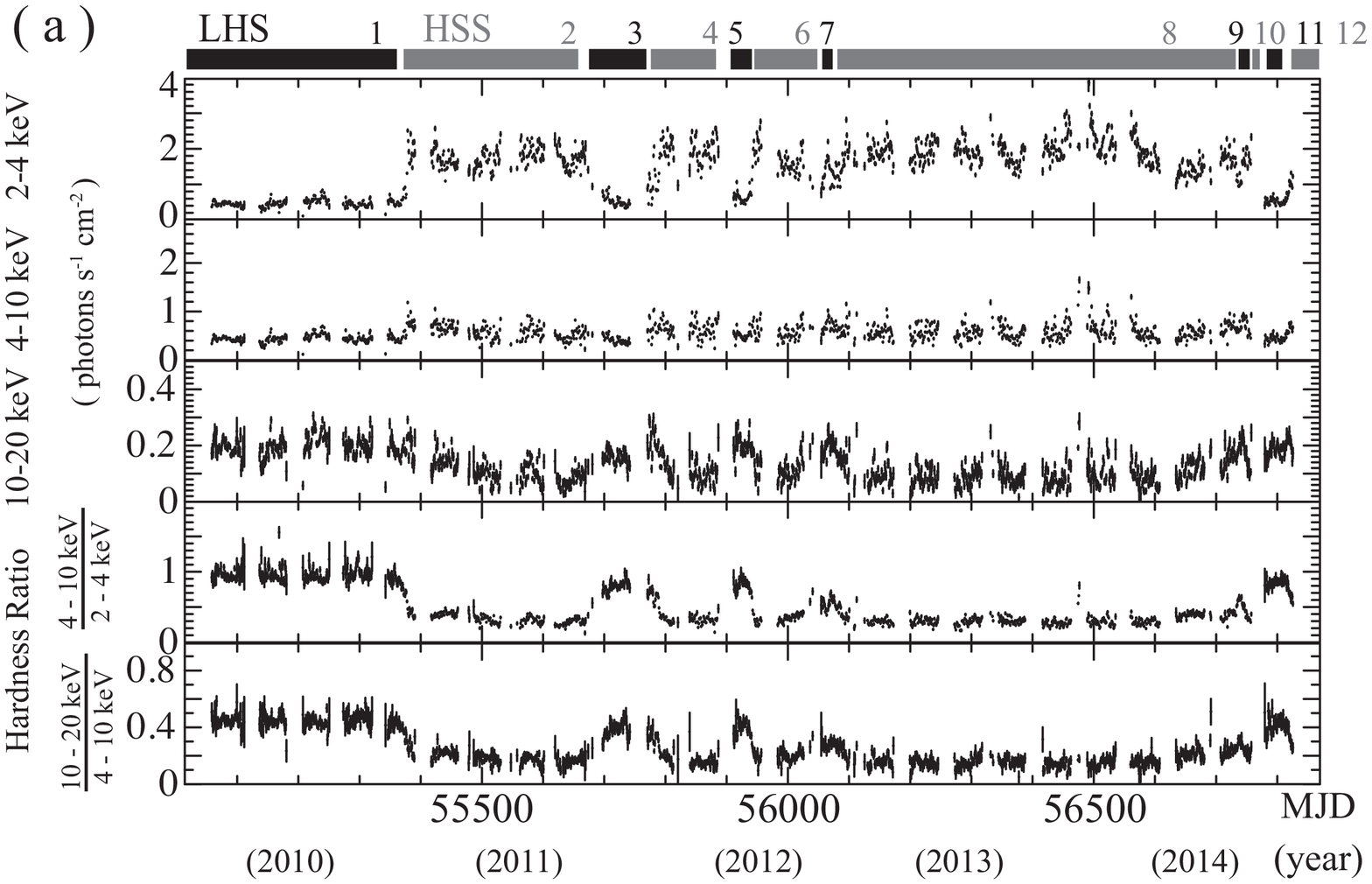}\\
   \vspace{1cm}
   \includegraphics[width=16cm]{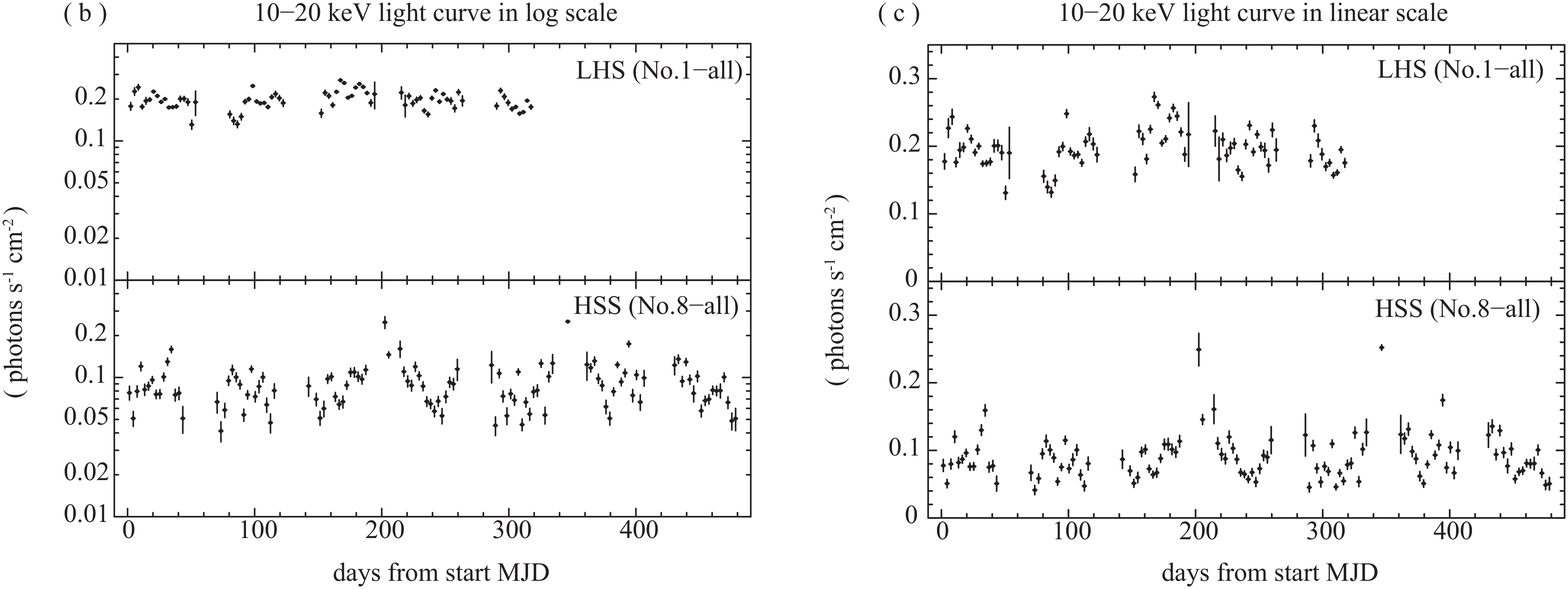}
  \end{center}
  \caption{(a) One-day bin light curves and HR histories of Cyg X-1 obtained with the MAXI/GSC.
  From the top to bottom panels, the 2--4 keV, 4--10 keV and 10--20 keV intensities,
  and the $I$(4--10 keV)/$I$(2--4 keV) and $I$(10--20 keV)/$I$(4--10 keV) ratios are plotted.
  The black and gray regions at the top indicate the LHS and the HSS periods, respectively.
  Data points with large error bars, due to high background counts, were omitted.
  (b) Expanded 10--20 keV light curves (three-day bin) of Data No.1-all and No.8-all,
  representing the LHS and the HSS, respectively.
  Ordinates are logarithmic.
  (c) The same as panel (b), but with linear scales.
  Data No.1-all and data No.8-all are shown LHS and HSS, respectively.
  The data No. is indicated in top panel, 
  and the corresponding MJDs are listed in table \ref{tb:hardem} and table \ref{tb:softem}.
  }
\label{fig:cygLC}
\end{figure*}

Figure \ref{fig:cygLC} shows one-day bin light curves of Cyg X-1 obtained with the GSC from 2009 August 15 (MJD = 55058) to 2014 November 9 (MJD = 56970),
in three energy bands (2--4 keV, 4--10 keV and 10--20 keV).
Time histories of two kinds of hardness ratios (HR), $I$(4--10 keV)/$I$(2--4 keV) and $I$(10--20 keV)/$I$(4--10 keV), are also plotted.
The state of Cyg X-1 can be recognized by the values of the HR.

Since the start of the MAXI observation in 2009 August, Cyg X-1 was in the LHS for about ten months.
At around MJD = 55378, the source made a transition to the HSS, where it stayed for another ten months.
After several times of state transitions, it has been mainly in the HSS since MJD = 56078.

\section{Analysis and results}
\subsection{Definition of spectral states}
\begin{figure}[h]
  \begin{center}
   \includegraphics[width=8cm]{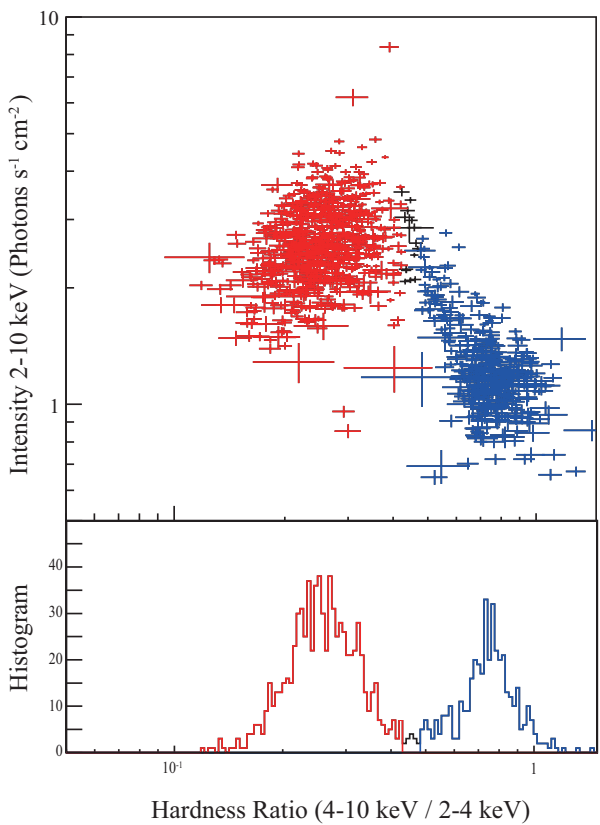}
  \end{center}
  \caption{
  The hardness-intensity diagram (top panel) and the histogram of the 4--10 keV vs. 2--4 keV HR (bottom panel).
  Blue and red data points specify the LHS and the HSS, respectively, while black points were taken during transitions.
  }
\label{fig:cygHDI}
\end{figure}

As shown in the HR histories in figure \ref{fig:cygLC}, Cyg X-1 exhibited distinct LHS and HSS.
Since the two HR histories have the same trend, we hereafter use the $I$(4--10 keV)/$I$(2--4 keV) ratio, which has better statistics. 
The top panel of figure \ref{fig:cygHDI} shows a hardness-intensity diagram.
Thus the two states are clearly separated.
The bottom panel in figure \ref{fig:cygHDI} shows a histogram of the HR, 
which exhibits two clear peaks corresponding to the HSS and the LHS.
We fit the histogram with two gaussian functions and determined their mean values and standard deviations.
Then we defined the boundaries of each state as 3$\sigma$ from the gaussian center, i.e.
HR$<$0.43 for the HSS and 0.48$<$HR for the LHS.

From figure \ref{fig:cygLC} and figure \ref{fig:cygHDI}, the periods of the two states have been obtained as listed in table \ref{tb:trans}.

\begin{table}
\small
\caption{Spectral states of Cyg X-1 during the MAXI observation.
Data numbers are indicated in figure \ref{fig:cygLC}.
}
\begin{center}
\begin{tabular}[b]{c c c c c c}
\hline
\hline
data         & spectrum & start & end & duration  \\ 
No.          & state    & MJD   & MJD & day       \\
\hline
1 & hard & 55058 & 55376 & 318 \\ 
2 & soft & 55378 & 55673 & 295 \\ 
3 & hard & 55680 & 55788 & 108 \\
4 & soft & 55789 & 55887 & 98  \\ 
5 & hard & 55912 & 55941 & 29  \\
6 & soft & 55943 & 56068 & 125 \\
7 & hard & 56069 & 56076 & 7   \\
8 & soft & 56078 & 56733 & 655 \\
9 & hard & 56735 & 56741 & 6   \\
10 & soft & 56742 & 56757 & 15 \\ 
11 & hard & 56781 & 56824 & 43 \\
12 & soft & 56854 & 56970$\sim$ & 116$\sim$ \\
\hline
\end{tabular}
\end{center}
\label{tb:trans}
\end{table}

\subsection{Power spectra}
\label{sec:psd}

\subsubsection{Definition of PSDs}

The X-ray light curve with 90-minute bin was converted to the PSD by the discrete Fourier transformation as
\begin{eqnarray}
F_{\rm c}(f) = \frac{2}{N}\sum_{j=1}^{N}y_{j}\cos(2\pi f\Delta t\times j) \nonumber \\
F_{\rm s}(f) = \frac{2}{N}\sum_{j=1}^{N}y_{j}\sin(2\pi f\Delta t\times j)
\label{eq:sincos}
\end{eqnarray}
where $y_{i}$ is the intensity (photons s$^{-1}$ cm$^{-2}$) at the $i$-th bin, $N$ is the number of data after filling the gaps as described later, 
and $\Delta t$ is a time interval of 5400 s, which is the regular sampling time of the MAXI public data, 
$f$ is the frequency which is an integer multiple of $\Delta f$=$\frac{1}{T}$, and  
$T$ is the total time span of the observation.
The factor of $2/N$ is employed so that $F_{c}(f)$ and $F_{s}(f)$ represent the amplitude of the cosine and sine components, respectively, regardless of $N$. 
The PSD is the sum of squares of $F_{\rm c}$ and $F_{\rm s}$ as, 
\begin{equation}
  P(f) = \frac{T}{2}\bigl\{F_{\rm c}^2(f) + F_{\rm s}^2(f)\bigr\}.
\label{eq:PSD}
\end{equation}
By multiplying with a factor of $T/2$, the derived PSD becomes independent of $T$, and has a unit of (RMS$^2$/Hz).
The PSD is normalized by the square of the average intensity following \citet{miyamoto1994}.
It is called normalized PSD (NPSD).

\subsubsection{Corrections for sampling and gap effects}
\label{psd_effect}

The MAXI data are not completely regular-sampled.
Sometimes data gaps are caused by sun avoidance, high particle background regions,
 small dead regions at the scan poles which move with the precession period of the orbit of the International Space Station, and other effects.
Therefore, we interpolated each gap with a linear line, 
which connects the pre-gap intensity (averaged over 5 data points) and the post-gap value (also averaged over 5 data points) \citep{sugimoto}.
Furthermore, the derived PSDs are strongly affected by aliasing effects, because MAXI measures the intensity of an X-ray source only for $40\sim 70$ s, every 5400 s period.
That is, each data point is a very short snapshot, with a long interval to the next (or from the preceding) sampling.
As a result, the MAXI light curves are generally sensitive to source variations not only on time scale longer than 5400 s, 
but also to these in between $40\sim 70$ s and $\sim 5400$ s.

In order to correct the PSDs for the effects of gaps (hereafter gap effects), and those of the short exposure with long-interval sampling (hereafter sampling effects), 
we simulated MAXI data of a variable source with specified variability.
Since the PSD of Cyg X-1 can be approximated by a power-law with an index of $-1$ (\cite{reig}; \cite{chura}),
this form of PSD was employed as the input variability.
Further assuming that the phases of the Fourier components are random, 
we produced a hundred fake light curves, each of which has 54 s time bin and covers the same time span as figure \ref{fig:cygLC}.
Details of the PSD simulation are given in Appendix \ref{gap}.
Then, for each simulation run, we picked up one data point from every 100 bins (= 5400 s = one orbit) and discarded the 99 points 
to simulate the sampling effects, and applied the same observing window as for Cyg X-1 in order to reproduce the gap effects.
Interpolating the gaps in the same way as for the actual data, a PSD was calculated from a fake light curve.
Finally, we took an average of the 100 simulated PSDs, and normalized it to the assumed input PSD, to obtain the transfer function 
(in Fourier space; figure \ref{fg:alias+gap} in Appendix \ref{gap}) 
of the present MAXI observation of Cyg X-1.
Below, PSDs obtained from the observed data are divided by the transfer function thus obtained, 
in order to correct for the gap effects and the sampling effects.
Prior to this division, we subtract the Poisson noise, 
which is estimated by a numerical simulation (Appendix \ref{poisson}).

\subsubsection{Results (1): using the longest data length}

For both the LHS and HSS, we calculated NPSDs from the light curves in the three energy bands, 2--4 keV, 4--10 keV and 10--20 keV.
In calculation, the longest span of data was used for each state, 
i.e. 318 days for the LHS 
and 477 days for the HSS.
These data segments are shown in table \ref{tb:hardem} (as No. 1-all) and table \ref{tb:softem} (as No. 8-all), respectively, together with other shorter segments available. 
Only the GSC data were used, because the SSC data are much more sparse than those of the GSC.
In the HSS, the selected span is shorter than Data No. 8 in table \ref{tb:trans} because short data bunches at the beginning and the end of the span were removed.
After subtracting the Poisson noise and correcting for 	the gap and the sampling effects (section \ref{psd_effect}), 
we obtained NPSDs down to 3$\times$10$^{-8}$ Hz as shown in figure \ref{fig:cygnpsd}.
The error of each data point was obtained by propagating errors of the observed light curve.
The NPSDs extend roughly with a power-law shape down to 3$\times$10$^{-8}$ Hz, 
but they scatter largely in the low frequency ($\leq 3\times10^{-7}$ Hz) region.

\subsubsection{Results (2): using shorter data lengths}
\label{psdresult2}

By shortening the data length to be used in a Fourier transform by equation (\ref{eq:sincos}), 
we can produce a larger number of NPSDs, and take their average. 
The NPSD obtained in this way is expected to suffer smaller statistical errors, 
although it lacks the information in the lowest frequencies.
Thus, the data span was restricted to 40 and 43 days for the LHS and the HSS, respectively, to cover a frequency range down to $\sim$ 3$\times$10$^{-7}$ Hz 
at the sacrifice of the $3\times$10$^{-8} - 3\times$10$^{-7}$ Hz range.
By doing so, the fraction of data gaps was reduced from $\sim$ 50 \% to $\sim$ 10\%. 
As listed in table \ref{tb:hardem}, six data segments were extracted for the LHS, and eight data segments in table \ref{tb:softem} for the HSS.
Then, we converted these data segments individually into NPSDs, and took their averages, separately over the LHS and HSS.
The obtained ensemble-averaged NPSDs are shown in figure \ref{fig:avepsd}, where the vertical error bars represent the NPSD scatter (standard deviation) within each ensemble.
In figure \ref{fig:avepsd}, ordinate employs power times frequency, 
instead of power itself.

\begin{figure*}
  \begin{center}
   \includegraphics[width=8cm]{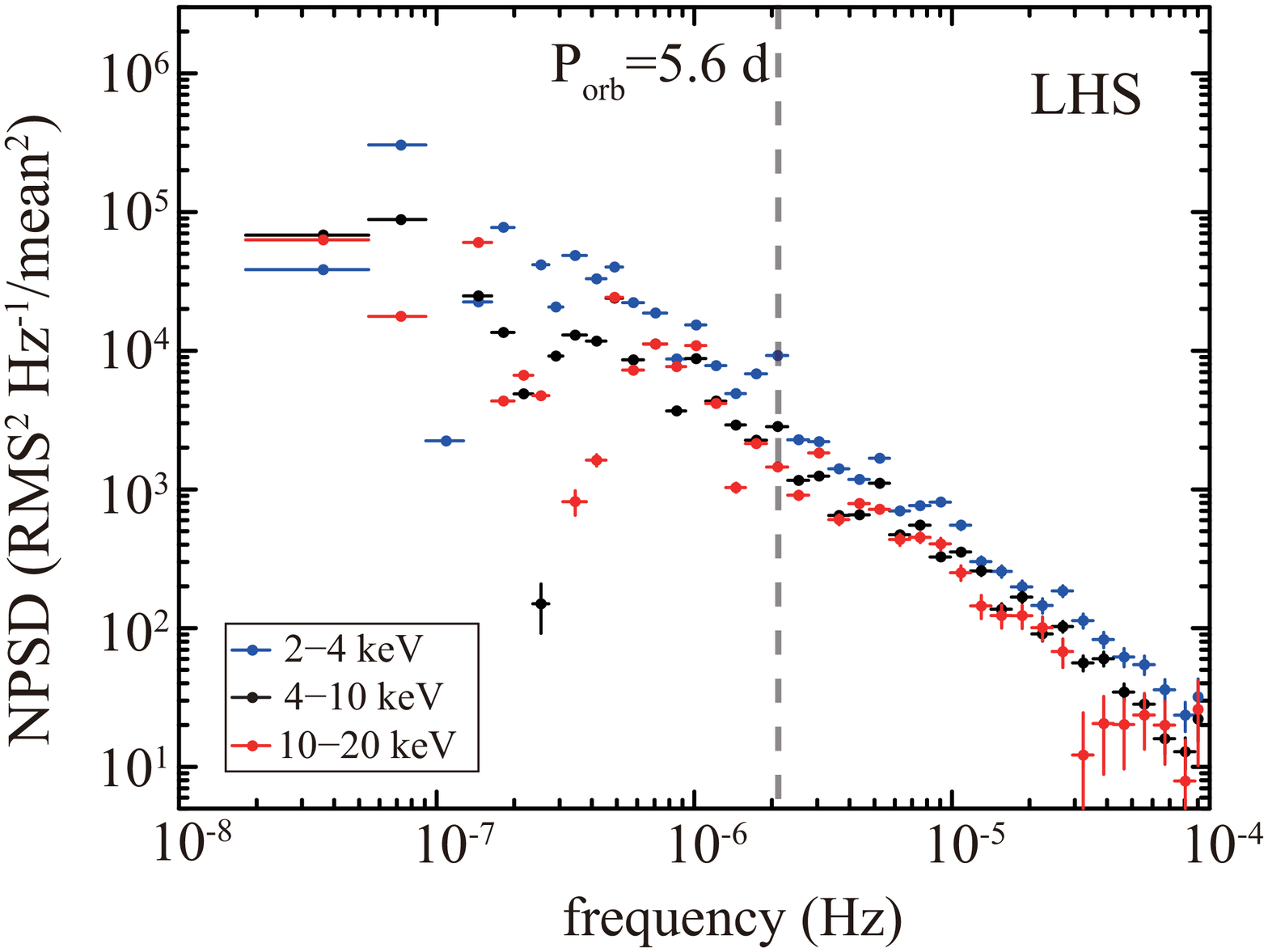}
   \hspace{5pt}
   \includegraphics[width=8cm]{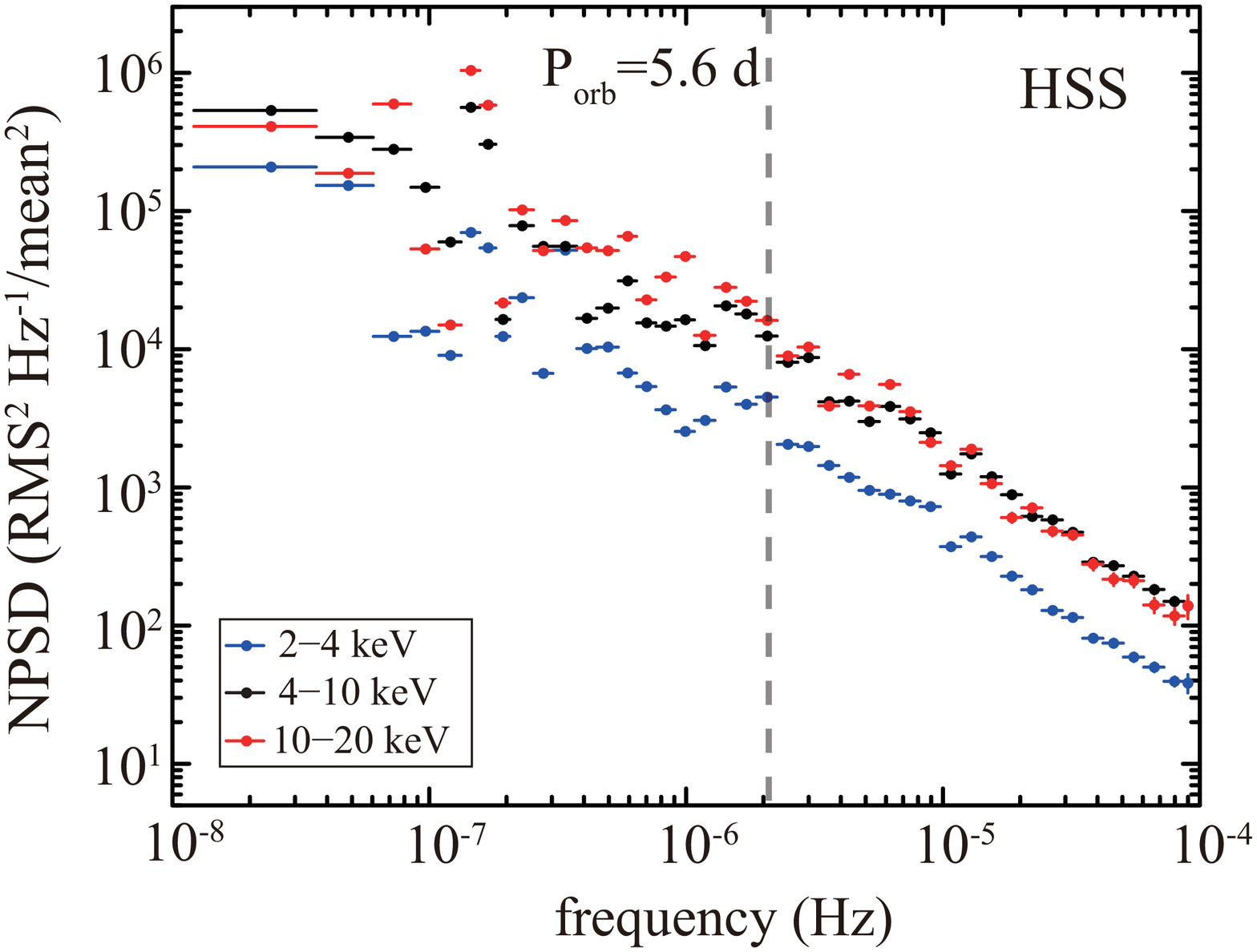}
  \end{center}
  \caption{NPSDs of Cyg X-1 in the LHS (left) and in the HSS (right), each calculated using the longest data span (No. 1-all in table \ref{tb:hardem} and No. 8-all in table \ref{tb:softem}).
  The blue, black and red points represent NPSDs in the 2--4 keV, 4--10 keV and 10--20 keV band, respectively.
  The vertical line at $2\times10^{-6}$ Hz indicates the 5.6-day orbital period.}
\label{fig:cygnpsd}
\end{figure*}

\begin{figure*}
  \begin{center}
   \includegraphics[width=8cm]{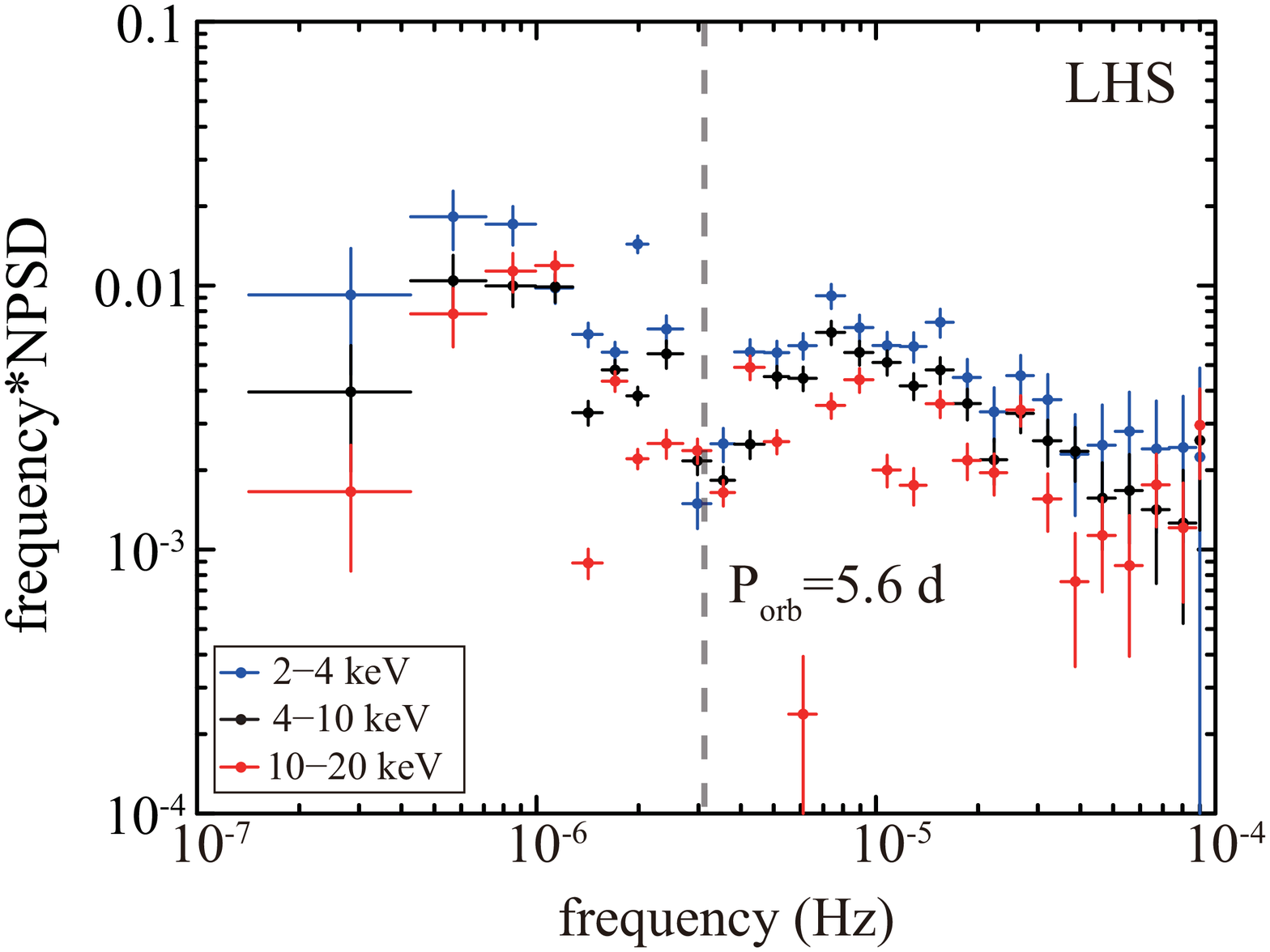}
   \hspace{5pt}
   \includegraphics[width=8cm]{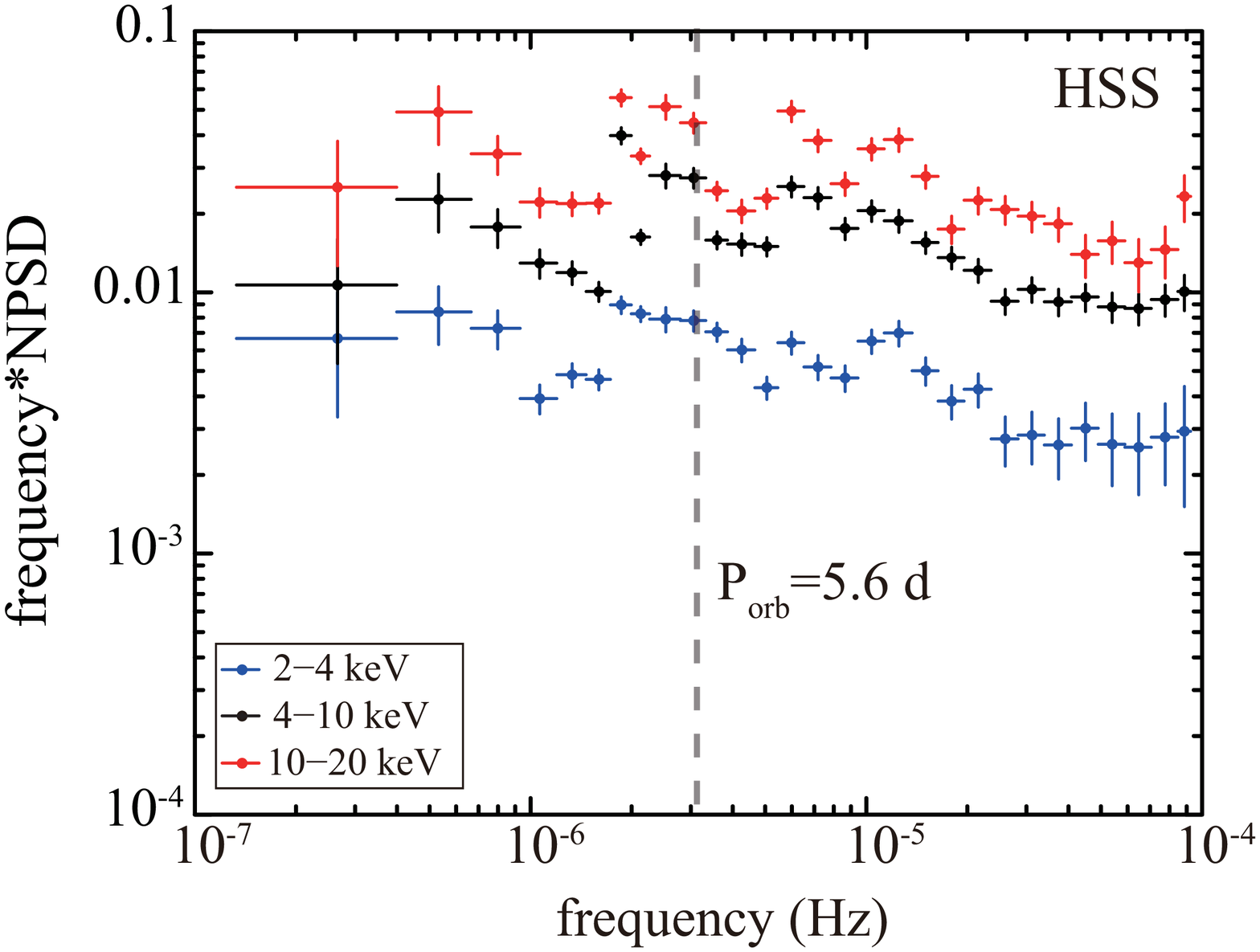}   
     \end{center}
  \caption{
  The NPSDs calculated over shorter data segments (40 d $\times$ 6 for the LHS and 43 d $\times$ 8 for the HSS), and averaged. 
  Here, the NPSDs are shown after multiplying with the frequency. 	
  }
\label{fig:avepsd}
\end{figure*}

As expected, the NPSDs in figure \ref{fig:avepsd} are less scattered even in the low frequency range
below $10^{-6}$ Hz, and show similar structures in all the energy bands, 
although they could still be subject to some structures including shallow dips around $3\times 10^{-6}$ Hz, 
and the weak tendency of flattening below $3\times 10^{-7}$ Hz.
As shown by a simulation in Appendix \ref{gap}, the possible flattening is unlikely to be caused by the data gaps.
Instead, it could be within the scatter of NPSDs, 
considering that the effect is seen in only the lowest-frequency data points in both states.	
We also note that the energy dependence in figure \ref{fig:cygnpsd} and figure \ref{fig:avepsd} is slightly different.
Especially, the normalization of the 4--10 keV NPSD in the HSS in figure \ref{fig:cygnpsd} is higher
than that in figure \ref{fig:avepsd}.
Considering the ensemble-averaging procedure involved in the NPSDs in figure \ref{fig:avepsd}, 
we regard them as better representing the Cyg X-1 variability than those in figure \ref{fig:cygnpsd}.

\begin{table}
\small
\caption{Data segments used in calculating NPSD in the LHSs.
}
\begin{center}
\begin{tabular}[b]{c c c}
\hline
data         & start & end \\ 
No.$^{*}$           & MJD   & MJD \\
\hline
1-all & 55058 &  55376  \\
\hline
1-1  &  55060  &  55100  \\ 
1-2  &  55138  &  55178  \\ 
1-3  &  55209  &  55249  \\
1-4  &  55277  &  55317  \\
3-1  &  55700  &  55740  \\
11-1  &  56781  &  56821  \\
\hline \\
\end{tabular}
\end{center}
* : ``$I$-$J$" denotes the $J$-th segment in the $I$-th data.
``$I$-all" means as the segments in the $I$-th data number.
\label{tb:hardem}
\end{table}

\begin{table}
\small
\caption{The same as table \ref{tb:hardem} but for the HSS.}
\begin{center}
\begin{tabular}[b]{c c c}
\hline
data         & start & end \\ 
No.       & MJD   & MJD \\
\hline
8-all & 56130 & 56607 \\
\hline
2-1  &  55487 & 55530  \\ 
2-2  &  55623 & 55666  \\ 
8-1  &  56201 & 56244  \\
8-2  &  56273 & 56316  \\
8-3  &  56418 & 56461  \\
8-4  &  56564 & 56607  \\
8-5  &  56637 & 56680  \\
12-1 &  56926 & 56969  \\
\hline
\end{tabular}
\end{center}
\label{tb:softem}
\end{table}

As seen in figure \ref{fig:avepsd},
the NPSD slope does not appear to depend significantly either on the spectral state or the energy band.
To confirm this suggestion, we fitted the individual NPDSs with a power-law, over a frequency range of $10^{-6}$ and $9.2\times10^{-5}$ Hz.
Then, the 18 NPSDs before taking ensemble averages (6 segments times 3 energy bands) in the LHS gave an average slope and the associated RMS scatter as $-1.35\pm 0.29$, 
whereas the 24 HSS ones (8 segments times 3 energy bands) gave $-1.29 \pm 0.23$.
Therefore, the NPSD slope is consistent with being independent of the state.
In addition, slope differences among the 3 energy bands are at most within the scatter of $\sim 0.3$ (in the LHS) and $\sim 0.2$ (in the HSS).

Figure \ref{fig:avepsd} reveals two more important properties. 
One is that the NPSD normalizations in the HSS are significantly energy dependent, increasing towards higher energies, 
whereas those in the LHS show a much weaker and opposite trend.
The other is that Cyg X-1 is much more variable, at least above 4 keV, on the relevant time scales while it is in the HSS than in the LHS.
To quantify these properties, 
we integrated the 6 (3 energy bands times two states) ensemble-averaged NPSDs from $10^{-6}$ to $9.2\times10^{-5}$ Hz,
and derived individually the fractional RMS variation $\eta$ as 
\begin{equation}
  \eta = \sqrt{2\Delta f \sum_{i=1}^{N_{F}}NPSD(f_i)}
\label{eq:rms}
\end{equation}
where $N_{F}$ is the number of data points in each NPSD.
The results, given in table \ref{tb:rms}, indeed confirm the above inferences
: the fractional RMS variation in 10--20 keV in the HSS ($\eta = 0.70$) is larger than that in the LHS ($\eta = 0.21$).
This property can be read directly from the light curve in a logarithmic scale presented in figure \ref{fig:cygLC}b. 
Meanwhile, when plotted in a linear scale (figure \ref{fig:cygLC}c), the absolute amplitude of variability is similar between the LHS and HSS.
These two properties can be explained as follows.
The 10--20 keV value of $\eta$ is $0.70/0.21\sim 3.3$ times higher in the HSS than in the LHS, 
while the average 10--20 keV intensities (as read from figure \ref{fig:cygLC}c) in the HSS is $0.089 / 0.19\sim 0.47$ times that in the LHS.
Therefore the absolute amplitude of variations in the HSS should be $3.3\times 0.47 = 1.55$ times that in the LHS.
This is close to unity, though not exactly the same.

\begin{table*}
\small
\caption{The fractional RMS $\eta$ of NPSDs from $10^{-6}$ to $9.2\times10^{-5}$ Hz.}
\begin{center}
\begin{tabular}[b]{c c c c c c c}
\hline
Spectral state        &                 &    LHS          &                  &                &    HSS          &               \\ 
\hline
Energy band (keV)     &       2-4       &      4-10       &      10-20      &        2-4      &       4-10      &     10-20      \\
\hline
$\eta$                &  0.31$\pm$0.09  &  0.24$\pm$0.05  &   0.21$\pm$0.04 &  0.31$\pm$0.09  & 0.53$\pm$0.12   &  0.70$\pm$0.29\\
\hline
\end{tabular}
\end{center}
\label{tb:rms}
\end{table*}

In the LHS, a peak corresponding to the orbital period of 5.60 d is 
clearly seen in the low energy band (2--4 keV) at $2.06\times$10$^{-6}$ Hz.
However, it is not clear in the other two energy bands in the LHS, 
in agreement with the Ginga/ASM results by \citet{kitamoto2000}.
In the HSS, the orbital period is not clearly seen in any energy band.

\begin{figure*}[htbp]
  \begin{center}
   \includegraphics[width=12cm]{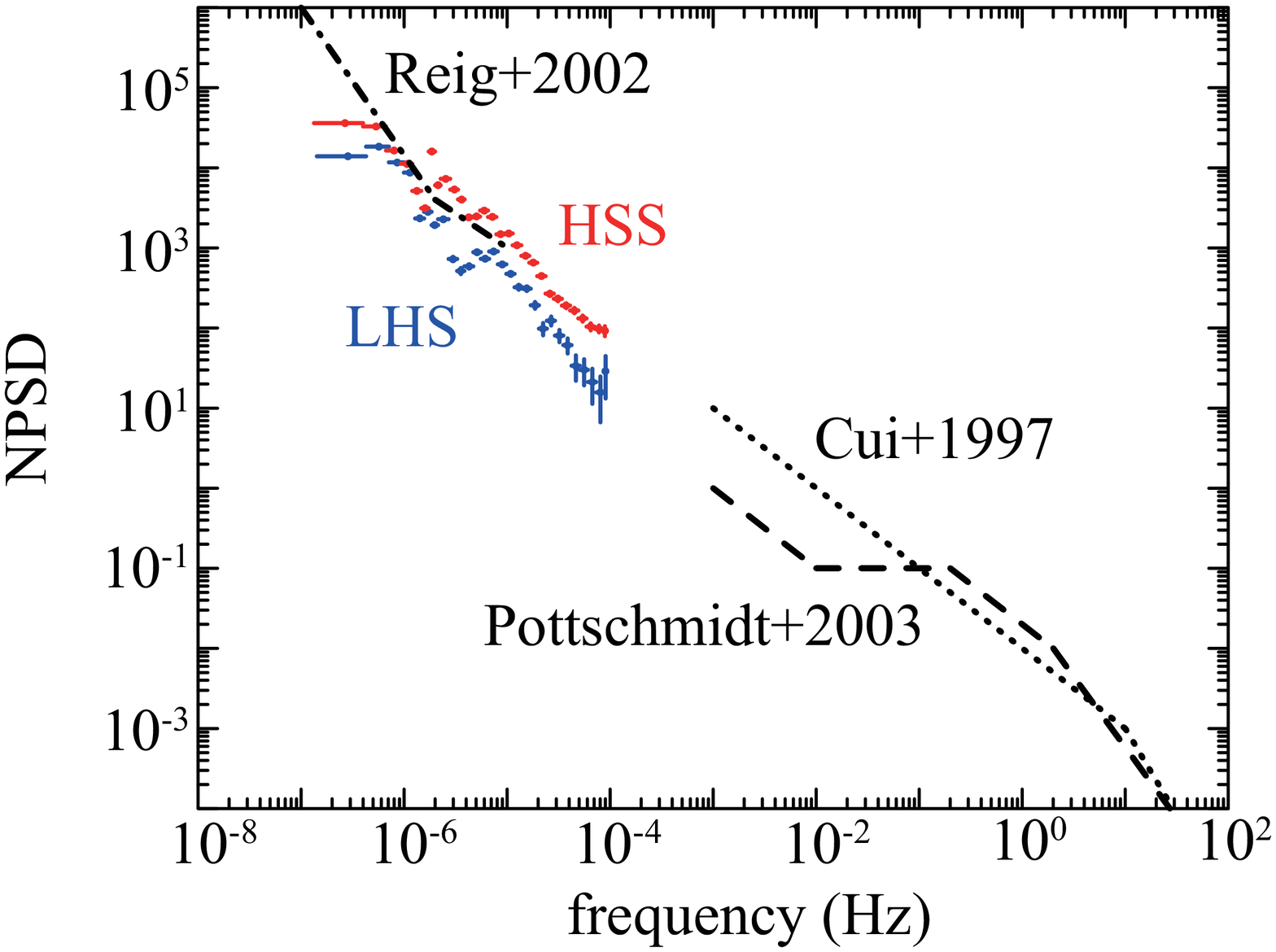}
  \end{center}
  \caption{A comparison of the MAXI 4--10 keV NPSDs in the LHS (blue) and the HSS (red), with the previous works.
  The dashed line is the 2--13 keV NPSD in the LHS by \citet{pot}, the dotted line is that in the HSS in 2--13 keV by \citet{cui1997}, 
  and the dot-dashed line is that by \citet{reig} where the most data are in the LHS.
  }
\label{fig:hkkchura}
\end{figure*}

Figure \ref{fig:hkkchura} shows our 4--10 keV NPSDs in both states, 
in comparison with previous works by \citet{cui1997}, \citet{pot}, and \citet{reig}.
In both states, our results at 10$^{-4}$ Hz locate on simple extrapolations of the NPSDs 
obtained previously by the RXTE/PCA in the frequency region above 10$^{-3}$ Hz.
The figure reconfirms and visualizes the higher variability in the HSS, 
already presented in table \ref{tb:rms}.
The NPSD in \citet{reig} is located between the two states of the present work.
This difference is presumably because their data include some of the HSS and the transition periods, 
although the main part is in the LHS.

\subsection{Energy Spectra}
\label{energy}

In order to investigate the spectral components that are responsible for the observed long-term variations,
we analyzed energy spectra of Cyg X-1 obtained with the MAXI/GSC and the SSC.
We used the same periods as in the PSD analysis,
i.e. from MJD = 55058 to MJD = 55376 for the LHS,
and from MJD = 56130 to MJD = 56607 for the HSS.
The SSC data, available only for about one third of the time, were also incorporated.

\begin{figure*}[hptb]
  \begin{center}
   \includegraphics[width=8cm]{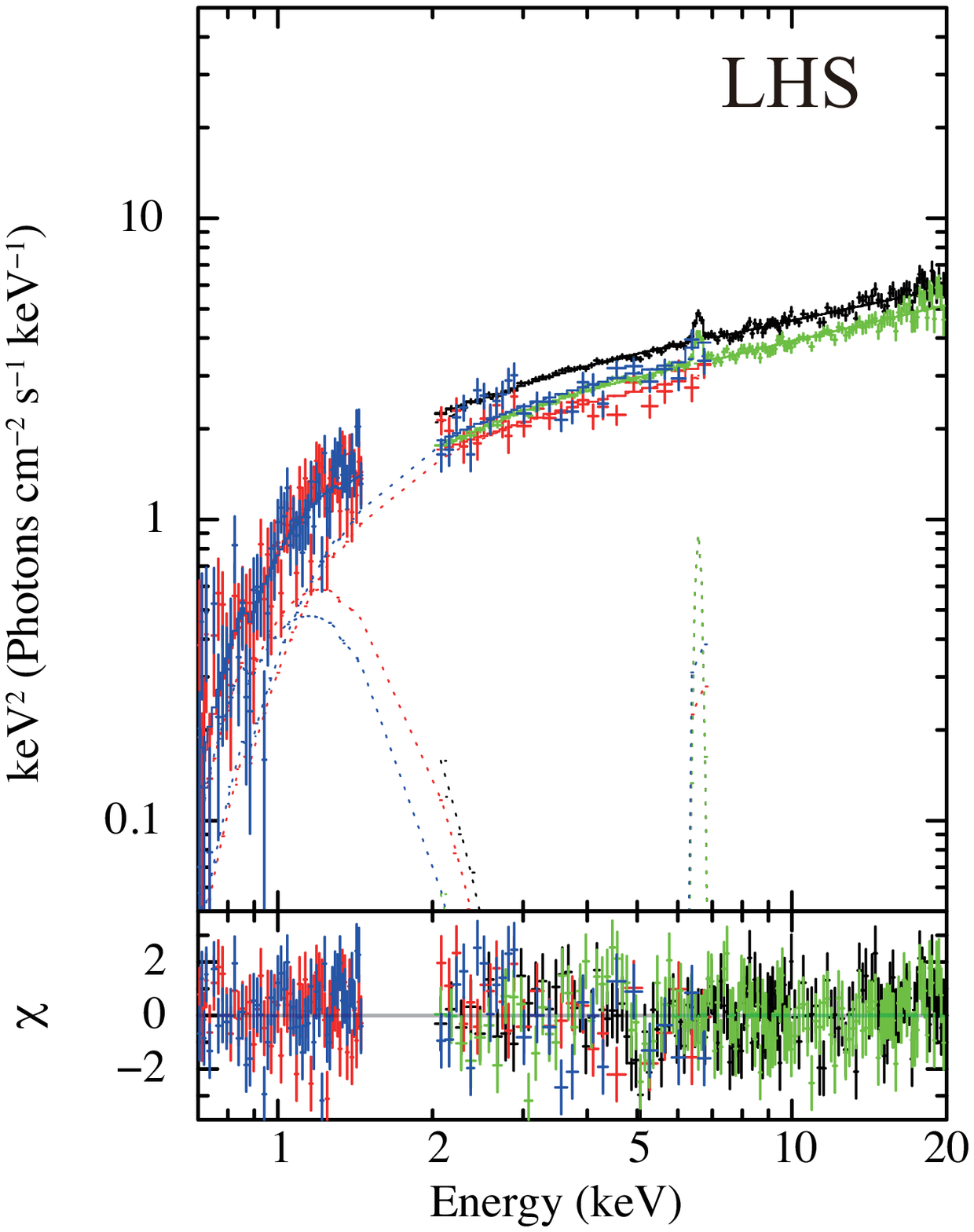}   
   \includegraphics[width=8cm]{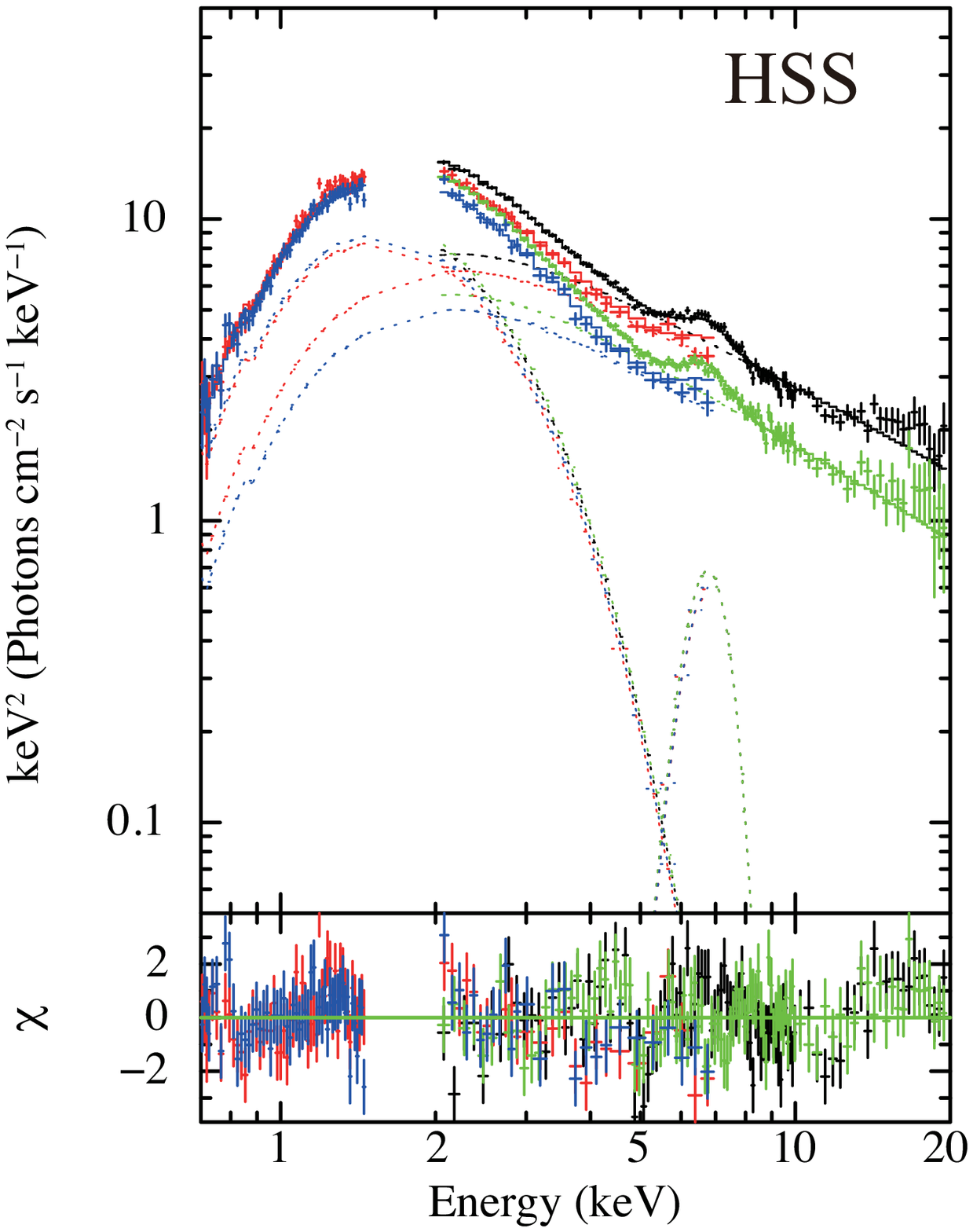}
  \end{center}
  \caption{Background-subtracted unfolded MAXI spectra in the LHS (left), and the HSS (right).
  Red and black indicate the bright-period data of the SSC and GSC respectively, 
  while blue and green indicate the faint periods of the SSC and GSC, respectively.
  The model is {\em phabs*(diskbb+nthComp+gaussian)}, 
  with the parameters listed in table \ref{tb:especcomp}.
  The 1.5--2 keV energy range is ignored in the fitting.
  The dotted lines represent contributions of the three components.
  The bottom panels are residuals from the model.}
\label{fig:cygespeccomp}
\end{figure*}

In section \ref{sec:psd}, we found that the long-term variability of Cyg X-1,
particularly in the HSS, is significantly energy dependent.
To reconfirm the implied spectral variation in this frequency range,
we divided the observed data into bright and faint periods
by comparing the individual one-day GSC intensities in 2--20 keV with their 15-day running averages.
Since the 15-day time scale corresponds to a frequency of $8\times 10^{-7}$ Hz, 
this procedure means an extraction of time variations longer than $8\times 10^{-7}$ Hz.
Then, eight spectra in total were produced;
the GSC and SSC data in the bright and faint periods, from the LHS and HSS.
The background-subtracted eight MAXI spectra obtained in this way are shown in figure \ref{fig:cygespeccomp}.
Thus, as expected, the HSS spectra are much softer than those of the LHS,
in good agreement with the general understanding of the spectral states.
For reference, these spectra include periods of increased low-energy absorption (seen as dips in the 2--4 keV light curves), 
which often appear for $\sim 15\%$ of the orbital phase.

\begin{figure*}[htbp]
  \begin{center}
   \includegraphics[width=16cm]{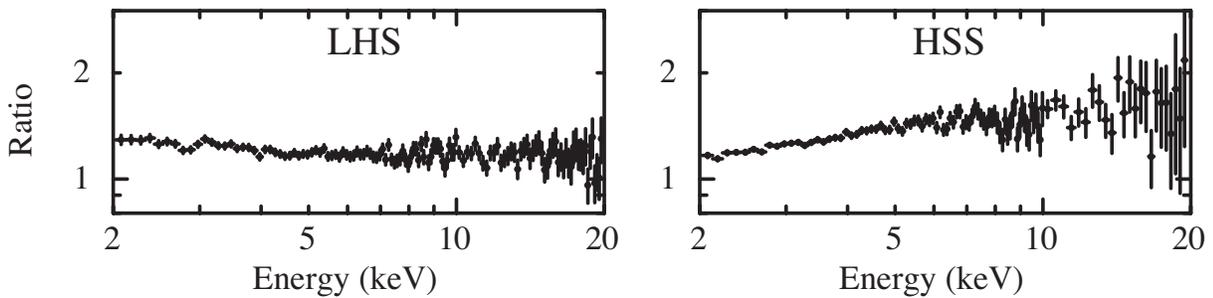}
  \end{center}
  \caption{Ratios of the GSC spectra in the bright period to that in the faint period,
  in the LHS (left panel), and in the HSS (right panel).
  }
\label{fig:pha}
\end{figure*}

To better visualize the intensity-correlated spectral changes in figure \ref{fig:cygespeccomp},
we present, in figure \ref{fig:pha},
the ratios of the GSC spectrum in the bright period to that in the faint period.
The SSC data are not shown here, because the errors are large.
In the LHS, the ratio decreases slightly with energy,
whereas the HSS ratio increases significantly towards higher energies.
To quantify these spectral results and examine their consistency with table \ref{tb:rms},
we  calculated the ratio $R$ of the photon flux in the bright period $F_{\rm b}$ to that in the faint period $F_{\rm f}$, and show the results in table \ref{tb:specrms}. 
We may express $R$ as
\begin{equation}
R = F_{\rm b}/F_{\rm f} = (\overline{F} +\Delta F)/(\overline{F}-\Delta F) ,
\label{eq:ratio}
\end{equation}
where $\overline{F} \equiv \frac{F_{\rm b}+F_{\rm f}}{2}$ is the average photon flux in the specified energy band
and $\Delta F \equiv \frac{F_{\rm b}-F_{\rm f}}{2}$ denotes the variable part.
In this formalism, the fractional RMS variation can be given as
\begin{equation}
 \eta' \equiv  a\Delta F/\overline{F} = a (R-1)/(R+1)
  \label{eq:eta}
\end{equation}
where $a$ is an appropriate numerical factor of order unity,
which can be approximated as energy independent.
This $a$ was introduced because the variation $\eta$ of equation (\ref{eq:rms}) calculated from NPSD can be different, by a constant factor, from that derived by equation (\ref{eq:eta}).
The derived values of $\eta'$ are also shown in table \ref{tb:specrms},
were $a=2.9$ has been adopted. 
Thus, the values of $\eta'$ agree, within errors,
with $\eta$ in table \ref{tb:rms} which is based on the PSD  analysis.

To understand the physical origin of the behavior seen in figure \ref{fig:pha},
as well as in table \ref{tb:rms} and table \ref{tb:specrms}, we proceeded to simultaneous model fitting to the 2--20 keV GSC spectrum
and the 0.7--7 keV SSC spectrum.
The 1.5--2 keV range of the SSC was excluded
to avoid the known systematic uncertainty in the effective area (\cite{tomida}).
First, we employed a model composed of a multi-color disk ({\em diskbb}) model (\cite{shakura, mitsuda}) for the low-energy part,
and a power-law ({\em powerlaw}) as a hard tail.
This model has often been used in the previous studies of BH X-ray binaries including Cyg X-1 (e.g. \cite{dotani}).
The photo-electric absorption model ({\em phabs}) with abundances by \citet{anders}
and the Fe-K emission line model ({\em gaussian}) were incorporated.
The model then becomes {\em phabs$\times$(diskbb+powerlaw+gaussian)} in the Xspec (\cite{arnaud96}) terminology; 
hereafter, we call it {\em diskbb+powerlaw} model.
The innermost radius of {\em diskbb} is derived as $r_{\rm{in}} = D\sqrt{N_{\rm{diskbb}}/\cos{i}}$,
where $N_{\rm{diskbb}}$ is the normalization of the {\em diskbb} model.
The distance $D$ from the earth and the inclination angle $i$ from our line of sight
were adopted to be 1.86 kpc and 27$^{\circ}$, respectively \citep{orosz}.
The hydrogen column density $N_{\rm H}$ and the gaussian parameters
were constrained to be common between the faint and the bright spectra  in each state,
but were allowed to differ between the two states.

The fit has been approximately successful,
and yielded the best-fit parameters as summarized in table \ref{tb:especpow}.
Thus, in both states, the  {\em diskbb} parameters are the same, within errors,
between the bright and faint spectra.
In the LHS, the {\em powerlaw} index is steeper in the bright spectrum than in the faint one,
while the opposite trend is found in  the HSS.
These are consistent with the spectral ratio shown in figure \ref{fig:pha}.
The $r_{\rm in}$ value for the LHS is consistent with that from a previous work \citep{yamada}.
On the other hand,  that for the HSS is smaller than that reported, e.g., in \citet{dotani} (even after corrections for the distance and the mass),
presumably because the luminosity in the present result
is roughly half  those in typical HSS observations made previously \citep{zhang}.

Referring to equation (\ref{eq:eta}), we calculated $\eta'$ also for the {\em powerlaw} component only, and show the results in table \ref{tb:specrms}. 
A comparison of the two values of $\eta'$, one from the entire photon counts while the other from the {\em powerlaw} contribution, 
reveals that the overall source variability in $> 4$ keV is determined solely by the power law variation, 
in agreement with the spectral decomposition in figure \ref{fig:cygespeccomp} (right).
In the 2--4 keV band, in contrast, the overall  $\eta'$ is reduced to $\sim 60\%$  of the {\em powerlaw} variability; 
this effect is readily attributed to the presence of the stable disk component.
In addition, table \ref{tb:specrms} reveals slight increase of  $\eta'$ (both total and {\em powerlaw}) from the 10-–20 keV to 4--10 keV bands.
The behavior of $\eta'$ agrees with that of $\eta$ in table \ref{tb:rms} and means that the {\em powerlaw} slope slightly hardens as it gets stronger, 
as visualized in figure \ref{fig:pha} (right) and quantified in the fitting results in table \ref{tb:especpow}.
This finding is not necessarily consistent, however,  with previous reports on Cyg X-1 in the HSS (e.g., \cite{jourdain}) and other HSS sources (e.g., \cite{koyama}),
that the hard-tail slope is relatively independent of the source intensity.
It is possible that the present HSS was somewhat atypical, 
because the values of $\Gamma \sim 2.9$ we measured is steeper than those ( $\Gamma \sim 2.4$; \cite{dotani}) observed from these BHBs in their typical HSS.

For a further analysis, we replaced the {\em powerlaw} with a more realistic model {\em nthComp} (\cite{zdziarski}), which represents Compton 
scattering of some soft photons by hot electrons ({\em diskbb+nthComp} model).
The seed photon source was set to be the {\em diskbb} component representing the disk emission.
Since the electron temperature $T_{\rm e}$ of the {\em nthComp} model cannot be determined with our data, we fixed it to a 
typical value of 100 keV \citet{Sunyaev79, makishima}. 
The best-fit parameters are summarized in table \ref{tb:especcomp} and the best-fit models are shown in figure \ref{fig:cygespeccomp}.
The fits are not yet formally acceptable, but the model, as shown in figure \ref{fig:cygespeccomp}, reasonably reproduces the spectra from 0.7 keV to 20 keV. 		
In this modeling, the disk photons are partially fed into {\em nthComp}, 
so that the parameter $r_{\rm in}$ in the {\em diskbb} model is smaller than the true inner radius which is denoted $R_{\rm in}$ hereafter. 
This $R_{\rm in}$ was calculated from the sum of 0.01--100 keV photon flux of the {\em diskbb} component, $F_{\rm{ph}}^{\rm{disk}}$, 
and that of nthComp component, $F_{\rm{ph}}^{\rm{nth}}$, utilizing the innermost temperature $T_{\rm in}$ of {\em diskbb} and the equation by \citet{kubota} as 
\begin{eqnarray}
\lefteqn{F_{\rm{ph}}^{\rm{disk}}+F_{\rm{ph}}^{\rm{nth}}} \nonumber \\
\lefteqn{= 0.0165\Biggl[\frac{R_{\rm{in}}^2\cos{i}}{(D/10\ \rm{kpc})^2}\Biggr]\Biggl(\frac{T_{\rm{in}}}{1\ \rm{keV}}\Biggr)^3\ \ \rm{ph}\ \rm{s}^{-1}\ \rm{cm}^{-2}.} \label{eq:r}
\end{eqnarray}
The derived values of $R_{\rm in}$, shown in table \ref{tb:especcomp}, are indeed larger than $r_{\rm in}$ in table \ref{tb:especpow} 
(although we do not discuss absolute values of $R_{\rm in}$).
The LHS is characterized by about twice larger $R_{\rm in}$ than the HSS, 
implying a disk truncation in the LHS as noted before \citep{makishima, yamada}.
Finally, in either state, the disk parameters ($T_{\rm in}$ and $R_{\rm in}$) are found to be apparently constant as the source varies.

\begin{table*}
\small
\caption{The flux ratio $R$ of equation (\ref{eq:ratio}), and the RMS variation $\eta'$ of equation (\ref{eq:eta}).}
\begin{center}
\begin{tabular}[b]{c c c c c c c}
\hline
Spectral state        &                 &    LHS          &                  &                &    HSS          &               \\ 
                      &           \multicolumn{3}{c}{$\hrulefill$}           &          \multicolumn{3}{c}{$\hrulefill$}         \\
\hline
Energy band (keV)     &       2-4       &       4-10      &       10-20      &       2-4      &       4-10      &       10-20      \\
\hline
Total                 &                 &                 &                  &                &                 &                 \\
~~~~~~~~~~~$R$        & 1.25$\pm$ 0.11  & 1.19$\pm$ 0.07  &  1.16$\pm$ 0.08  & 1.17$\pm$ 0.20 & 1.43$\pm$ 0.24  &  1.64$\pm$ 0.31  \\
~~~~~~~~~~~$\eta'$    & 0.32$\pm$ 0.10  & 0.25$\pm$ 0.06  &  0.21$\pm$ 0.08  & 0.23$\pm$ 0.19 & 0.52$\pm$ 0.20  &  0.70$\pm$ 0.23  \\
Powerlaw              &                 &                 &                  &                &                 &                 \\
~~~~~~~~~~~$R$        & 1.29$\pm$ 0.12  & 1.19$\pm$ 0.07  &  1.16$\pm$ 0.08  & 1.33$\pm$ 0.16 & 1.47$\pm$ 0.22  &  1.64$\pm$ 0.31 \\
~~~~~~~~~~~$\eta'$    & 0.37$\pm$ 0.11  & 0.25$\pm$ 0.06  &  0.21$\pm$ 0.08  & 0.41$\pm$ 0.14 & 0.55$\pm$ 0.18  &  0.70$\pm$ 0.23 \\
\hline
\end{tabular}
\end{center}
\label{tb:specrms}
\end{table*}

\begin{table*}[htpb]
\small
\caption{The best-fit parameters of the absorbed {\em diskbb} plus {\em powerlaw} model.}
\begin{center}
\begin{tabular}[b]{cccccc}
\hline \hline
\multicolumn{6}{c}{Model = phabs*(diskbb+powerlaw+gaussian)}\\ \hline
Component     & Parameter                    & LHS/bright               & LHS/faint                    & HSS/bright              & HSS/faint     \\ \hline
phabs         & $N_{\rm H}^{*}$              & \multicolumn{2}{c}{$5.8\pm 0.6$}                      & \multicolumn{2}{c}{$4.92\pm 0.12$}     \\ \hline
diskbb        & $T_{\rm{in}}$ (keV)          & $0.24\pm 0.02$           & $0.21\pm 0.02$             & $0.50\pm 0.01$          & $0.49\pm 0.01$ \\ 
              & $r_{\rm{in}}^{\dagger}$ (km) & $84^{+34}_{-22}$         & $99^{+49}_{-26}$           & $33^{+1}_{-5}$          & $36^{+1}_{-5}$   \\ \hline
powerlaw      & Index                        & $1.66\pm 0.01$           & $1.62\pm 0.01$             & $2.85\pm 0.03$          & $2.98\pm 0.03$ \\
              & Norm$^{\ddagger}$                         & $2.09\pm 0.04$           & $1.65^{+0.04}_{-0.03}$     & $19.5^{+1.0}_{-1.0}$  & $16.7\pm 1.0$    \\ \hline
gaussian      & Line E (keV)                 & \multicolumn{2}{c}{$6.66^{+0.17}_{-0.16}$}            & \multicolumn{2}{c}{$6.62\pm 0.07$}  \\ 
              & Sigma (keV)                  & \multicolumn{2}{c}{$0.7$\ (fixed)}                    & \multicolumn{2}{c}{$0.70^{+0.09}_{-0.08}$}  \\ 
              & Norm$^{\S}$            & \multicolumn{2}{c}{$7.2\pm 1.6$}                      & \multicolumn{2}{c}{$28.3^{+3.7}_{-3.2}$}  \\ \hline        
fit\ goodness & $\chi ^{2}_{\nu}(\nu)$       & \multicolumn{2}{c}{1.38 (420)}                       & \multicolumn{2}{c}{1.31 (356)}    \\ 
\hline
\end{tabular}
\end{center}
* : In a unit of $10^{21}$ cm$^{-2}$.\\
$\dagger$ : The distance is assumed to be 1.86 kpc and the inclination angle is assumed to be 27$^{\circ}$ \citep{orosz}.\\
$\ddagger$ : In a unit of photons cm$^{-2}$ s$^{-1}$ keV$^{-1}$ at 1 keV.\\
$\S$ : In a unit of 10$^{-3}$ photons cm$^{-2}$ s$^{-1}$.\\
\label{tb:especpow}
\end{table*}
\begin{table*}[htpb]
\small
\caption{The best-fit parameters of the model consisting of a {\em diskbb} and {\em nthComp} taking account of the photo-electric absorption.}
\begin{center}
\begin{tabular}[b]{cccccc}
\hline \hline
\multicolumn{6}{c}{Model = phabs*(diskbb+nthComp+gaussian)}\\ \hline
Component        & Parameter                                     & LHS/bright              & LHS/faint                     & HSS/bright              & HSS/faint       \\ \hline
phabs            & $N_{\rm H}^{*}$                               & \multicolumn{2}{c}{$6.1\pm 0.6$}                      & \multicolumn{2}{c}{$3.6\pm 0.1$}  \\ \hline
diskbb           & $T_{\rm{in}}$ (keV)                           & $0.23\pm 0.02$           & $0.20\pm 0.02$             & $0.48\pm 0.01$          & $0.47\pm 0.01$    \\ 
                 & $F_{\rm{ph}}^{\rm{disk}\dagger}$       &  47.0                    & 43.2                       & 69.6                    & 72.5   \\ \hline
nthComp$^{\ddagger}$   & Gamma                                         & $1.68\pm 0.01$           & $1.65\pm 0.01$             & $2.81\pm 0.03$          & $2.91\pm 0.04$       \\
                 & $F_{\rm{ph}}^{\rm{nth}\dagger}$        & 12.9                     & 10.9                       & 36.3                    & 27.4    \\ \hline
gaussian         & LineE (keV)                  & \multicolumn{2}{c}{$6.65\pm 0.16$}                    & \multicolumn{2}{c}{$6.64\pm 0.07$}   \\ 
                 & Sigma (keV)                  & \multicolumn{2}{c}{$0.7$\ (fixed)}                    & \multicolumn{2}{c}{$0.64^{+0.09}_{-0.08}$}  \\ 
                 & Norm$^{\S}$            & \multicolumn{2}{c}{$7.6^{+1.6}_{-1.5}$}               & \multicolumn{2}{c}{$24.5^{+3.1}_{-2.8}$}   \\ \hline
inner radius     & $R_{\rm{in}}^{l}$ (km)         & $116\pm15$  & $127\pm21$                    & $54\pm 2$    & $53\pm 2$ \\ \hline
Luminosity$^{\#}$&                       & $1.76\pm 0.04$           & $1.50\pm 0.03$             & $2.71\pm 0.02$          &$2.35\pm 0.02$ \\ \hline          
Compton fraction$^{**}$ &                       & 0.87                     & 0.89                       & 0.51                    & 0.41 \\ \hline   
fit\ goodness    & $\chi ^{2}_{\nu}(\nu)$       & \multicolumn{2}{c}{1.37 (420)}                        & \multicolumn{2}{c}{1.38 (370)} \\ 
\hline
\end{tabular}
\end{center}
$*$ : In a unit of $10^{21}$ cm$^{-2}$.\\
$\dagger$: Photon flux in a unit of photons cm$^{-2}$ s$^{-1}$ in the energy range of 0.01-100 keV.\\
$\ddagger$ : The electron temperature $T_{\rm{e}}$ is fixed at 100 keV. The spectrum of the seed photon is diskbb and the temperature $T_{\rm{bb}}$ is fixed at $T_{\rm{in}}$.\\
$\S$ : In a unit of $10^{-3}$ photons cm$^{-2}$ s$^{-1}$.\\
$l$: The distance is assumed to be 1.86 kpc and the inclination angle is assumed to be 27$^{\circ}$ \citep{orosz}.\\
$\#$ : In a unit of 10$^{37}$ erg s$^{-1}$ and in the energy range of 0.5-100 keV.\\
$**$ : The fractional luminosity in {\em nthComp}, which is calculated from the disk luminosity $L_{\rm{disk}}$ and the Compton luminosity $L_{\rm{nth}}$ as 
$\frac{L_{\rm{nth}}}{L_{\rm{disk}}+L_{\rm{nth}}}$.\\
\label{tb:especcomp}
\end{table*}

\section{Discussion}

\subsection{Summary of data analysis}

Using 5 years archival data obtained with MAXI, we derived the NPSDs of Cyg X-1 in its LHS and HSS from $10^{-7}$ Hz to 10$^{-4}$ Hz in the three energy bands.
It is of particular importance that the long-term variations in the two states were studied in a unified way 
using the same instrument, with the same analysis method, 
and under similar data statistics.
In addition, the long-term NPSD in the HSS was obtained for the first time thanks to a fortunate opportunity that Cyg X-1 stayed in the HSS in most of the time since 2010 June.
These results on the HSS are expected to provide some clues to the still unknown origin of the hard-tail component, 
which is nearly always observed in this spectral state.

In the LHS, the NPSD down to $10^{-7}$ Hz obeys a power-law, and  is approximately expressed as an extrapolation of the NPSD above 0.01 Hz \citep{pot, nowak}.
We found that the NPSD in the low (2--4 keV) energy band is slightly larger than those in the 4--10 and 10--20 keV bands 
(figure \ref{fig:cygnpsd} left, figure \ref{fig:avepsd} left, and table \ref{tb:rms}).

The newly obtained $10^{-7}-10^{-4}$ Hz NPSD in the HSS is also approximated by a power-law, 
and on an extrapolation of the NPSD previously obtained in frequencies above 10$^{-4}$ Hz \citep{chura, cui1997}.
The NPSD (RMS$^2$/mean$^2$ Hz$^{-1}$) in the HSS is about one order of magnitude larger than that in the LHS in 10--20 keV.  
This differs from the case of other BH binaries reported by \citet{miyamoto1993}, although their data were limited to higher frequency range above 0.01 Hz.
In the HSS, the NPSD from $10^{-7}$ Hz to 10$^{-4}$ Hz has an energy dependence in such a way that it is 5$\sim$6 times higher in the 10--20 keV band than that in the low energy band (2--4 keV).  
This finding extends the results of energy dependence in 0.2--200 Hz reported by \citet{grinberg}. 
These energy dependences of the long-term variation in the HSS were reconfirmed via the spectral analysis in section \ref{energy}.

In the present paper,  
we used only the MAXI data 
because the energy band is suitable for simultaneous analyzing the variations of the disk component and the power-law component.
In order to further study the hard tail variability, 
the Swift/BAT data with good statistic above 20 keV should be utilized: 
we consider that this is our future task.

\subsection{Comparison with short-term variability}
Since the first discovery by \citet{oda}, the aperiodic fast X-ray variation of Cyg X-1 has been
investigated by a number of authors (e.g., \cite{negoro1994}, \cite{chura}),
mainly in the LHS and typically over $10^{-3}-10^2$ Hz frequency range.
Similar studies have been performed on other BHBs as well, including in particular GX 339-4 (\cite{maejima}; \cite{miyamoto1994}).
Given these, let us briefly compare the present studies with the previous results on the short-term variability of Cyg X-1.

Our results on the LHS has two similarities to the short-term variability in the same state.
One is the PSD slope, which is $\sim 1.4$ over $10^{-7}-10^{-2}$ Hz and $\sim 1$ in the $>10^{-1}$ Hz range (figure \ref{fig:hkkchura}).
The other is the energy dependence; on both time scales, $\eta$ decreases slowly towards higher energies, 
implying ``softer when brighter" characteristic, 
which Cyg X-1 shows when it is above $\sim 1\%$ of the Eddington luminosity.
For example, \citet{grinberg} reported that the 0.125--256 Hz variation of Cyg X-1 in the LHS is about twice higher in 2.1--4.5 keV than in 5.7--9.4 keV,
in approximate agreement with our table \ref{tb:rms}.
Also, the Suzaku result obtained in the LHS,
namely, figure 8b of \citet{makishima},
reveals a similar energy dependence.
These two resemblances altogether suggest a common mechanism working on these very wide range of time scales.
For example, the higher short-term variability in softer energy bands may be a result of spectral softening as individual ``shots" develop 
\citep{negoro1994, yamadashot}.
However, the phenomenon cannot be completely self-similar, since the PSD shows a considerable flattening over the $10^{-2}-10^{-1}$ range.

The fast variability in the HSS has been considerably less studied than that in the LHS.
Nevertheless, 
we can point out again two similarities between the long-term vs. short-term variations.
One is that our NPSD index in the HSS ($\sim 1.3$) is similar to the short-term HSS index of $\sim 1$ by \citet{cui1997}, 
as clearly seen figure \ref{fig:hkkchura}.
The other is the energy dependence; the slight increase of variability from the 4--10 keV to 10--20 keV ranges 
(subsection \ref{psdresult2} and section \ref{energy}) 
is also observed in short-term results by \citet{grinberg}.	
As already described, this ``harder when brighter" property is opposite to the behavior, ``softer when brighter", in the LHS.

\subsection{Spectral variation}

The energy spectrum in the LHS is known to be approximated by a power-law with a high-energy cutoff around 100 keV (e.g., \cite{makishima}).  
This is interpreted by a scenario that an optically-thick standard disk \citep{shakura} is truncated at some distance 
from the BH and an optically-thin accretion flow (or a corona) is formed around the BH \citep{esin}.  
High temperature electrons, in the optically-thin corona, up-scatter seed photons presumably from the outer standard disk.
In this frame work, the fast hard X-ray variability in the LHS may be explained by considering 
that the corona covers a variable fraction of the disk \citep{makishima}; 
when this fraction increases, 
the source gets brighter, 
and softer due to the enhanced Compton cooling of the corona.
In fact, the MAXI spectrum accumulated over the LHS (figure \ref{fig:cygespeccomp} left) was described successfully with the two component model, 
consisting of a low-temperature {\em diskbb} \citep{mitsuda} representing the emission from such a truncated standard disk, 
and the {\em powerlaw}  or the {\em nthComp} representing the Compton up-scattered component.

Like in many other reports on BHBs in the HSS, the MAXI spectrum in the HSS has been expressed by a dominant optically-thick thermal spectrum, 
accompanied by a {\em powerlaw} component extending into higher energies.
The former is again interpreted by an emission from a standard-disk \citep{shakura}, which is now considered to extend down to 
the innermost stable circular orbit (ISCO) around the BH.  
This is supported by the fact that the obtained inner disk radii in the HSS do not differ significantly between the faint and bright periods.
Like the overall hard X-ray emission in the LHS, 
the hard {\em powerlaw} component in the HSS could also be a Compton up-scatters component produced by some hot electrons 
\citep{cui1998, gierlinski}, but details are still unclear.

In order to better understand the NPSD results, we fitted the spectra by a model composed of a {\em diskbb}, and the {\em powerlaw} or the {\em nthComp}, 
and studied how the spectrum changes as the source varies on frequencies range below $8\times 10^{-7}$ Hz.
The obtained results, fully consistent with those from the NPSD studies, can be summarized into the following four points.
\begin{enumerate}
\item[1] In  both states, the {\em powerlaw} component is responsible for the observed long-term intensity variations, while the disk emission is essentially constant.
\item[2] The power-law component in the HSS is concluded to be more variable, on this frequency range, than that in the LHS.
\item[3] The decreasing variability towards higher energies, observed in the LHS, can be attributed mostly to the ``softer when brighter" property of the {\em powerlaw} component.
\item[4] The variability increase with energies in the HSS results from a combination of the two effects,
namely, the ``harder when brighter" property of the {\em powerlaw} component in this state, and the presence of the stable disk emission at lower energies.
\end{enumerate}

Similarly high long-term variations were observed also from some BH transients, 
including GS 1124$-$684 and GS 2000+25 \citep{tanaka, terada}.
When their outburst decline was followed by sparse snap-shot observation with Ginga, the intensity of the disk component changed smoothly, 
but the power-law component varied largely from one observation to the next.
Thus, high long-term variability of the hard component in the HSS may be common to all BH binaries, 
including persistent and transient sources.


\subsection{Origin of the time variation}
Let us discuss the origin of the long-term variations which we detected down to $10^{-7}$ Hz.
The simplest possibility would be that  fluctuations in
the overall mass transfer rate from the companion star
produce the observed long-term X-ray variations.
Generally, the mass transfer in Cyg X-1 is thought to
occur via capture of the stellar winds from the super giant companion star, HDE226868.
The winds are considered inhomogeneous as reported by several authors \citep{Conti89,GiesBolton86,Hanke09},
and as evidenced by variable X-ray absorption episodes
around the superior conjunctions of the BH \citep{kitamoto, remillard1984, church1997}.
Therefore, the mass accretion rate $\dot{M}$ must fluctuate,
and could produce the long-term variability.
However, besides the issue of how these fluctuations can propagate
to the X-ray emission regions (see below),
it is not trivial to explain with this scenario
the difference in the variation ($\eta$ and $\eta'$)
observed between the LHS and HSS.

The observed long-term X-ray variations may alternatively arise
when some  fluctuations in $\dot{M}$,
produced spontaneously in the accretion flow at various radii,
propagate to the BH vicinity.
At a radius $r$ from the BH,
such spontaneous fluctuations
may generally take place
on a time scale comparable to the thermal time scale
$t_{\rm th}=(\alpha \Omega_{\rm K})^{-1}$,
or the dynamical time scale $t_{\rm d} = (\Omega_{\rm K})^{-1}$,
where $\Omega_{\rm K}$ is the Keplerian angular velocity at $r$
and $\alpha$ is the viscosity parameter.
Such variations can propagate toward the center
only if their time scales are longer than the viscous time scale,
given as $t_{\rm v}=(\alpha \Omega_{\rm K})^{-1}(r/H)^{2}$,
where $H$ is the disk half thickness.
In an optically thick and geometrically thin  disk,
this $t_{\rm v}$ is much longer than $t_{\rm th}$ and $t_{\rm d}$ because $H \ll r$.
Therefore, variations in $\dot{M}$ produced at any outer radii,
 being strongly dissipated, would not propagate down to the X-ray emitting region.
However, if some fraction of $\dot{M}$ streams through
a geometrically-thick ($H\sim r$) and optically-thin flow,
in which  we expect $t_{\rm v} \sim t_{\rm th}$,
the fluctuations produced at various larger radii can propagate inwards,
and cause the X-ray intensity to vary on a wide range of time scales.
This idea was invoked by \citet{chura} to explain
the X-ray variations on  timescales down to $10^{-4}$ Hz.
We may extend this two-flow scenario from the time scale of $10^{-4}$ Hz to $10^{-7}$ Hz,
identifying the optically-thick part with the standard accretion disk,
and assuming that the optically-thin part
develops into the Comptonizing corona in a vicinity of the BH.
Assuming a $15~M_{\odot}$ BH and $\alpha = 0.01$,
the observed time scale of $t_{\rm th} \sim 10^{7}$ s
can be explained if the optically-thin flow starts
at $\sim 3 \times 10^{12}$ cm ($\sim 6 \times 10^{5} R_{\rm s}$).
Furthermore, as detailed in Appendix \ref{mdot},
the larger long-term variability in the HSS can be explained
if  the fraction of $\dot{M}$ through the optically-thin part
carries $\sim 0.5~\dot{M}$ in the HSS,
while $\sim 0.2~\dot{M}$ in the LHS.

Although the two-flow picture, as constructed above after \citet{chura},
can account for essential features of the observed long-term variability of Cyg X-1,
it is still subject to a few issues to be solved.
One is that the required size of the optically-thin flow is comparable
to the binary size of Cyg X-1 (0.2 AU).
It is not obvious whether such a large-scale optically-thin flow
can be actually created by possible candidate mechanisms,
such as X-ray irradiation, vertical magnetic pressure,
and initial scatter in the specific angular momentum of accepting blobs.
Even if such a flow is produced,
it is not obvious whether its low density can be compensated by its higher radial velocity,
to  carry a considerable fraction of the total $\dot{M}$.
Furthermore, we need to assume that the state difference,
which is usually thought to be triggered at close vicinities of the BH,
is already present at such large radii:
it is not clear, either,
whether some feedback mechanisms (e.g., X-ray irradiation)
can control large-scale accretion flows in the required manner.

At present, we cannot conclude for sure 
whether the observed slow variations in the HSS can be explained by either the stellar wind fluctuation or the two-flow picture.
We might need to consider other origins of the slow variability,
including their production at regions much closer to the BH.
Further discussion is beyond the scope of the present paper:
although our ultimate goal is to identify the origin of the long-term variation, 
the currently available data information is still insufficient for that purpose.

\section{Conclusion}
By  analyzing the three-band (2--4, 4--10 and 10--20 keV) MAXI data accumulated over 5 years,
we studied characteristics of the long term X-ray variation of Cyg X-1
over the frequency range of $10^{-7}-10^{-4}$ Hz.
The long-term NPSD in the HSS was obtained for the first time.
By treating the LHS and the HSS data separately 
but in the same manner, 
the following results have been obtained.
\begin{enumerate}
\item 
In the LHS and HSS, the index of the NPSD was obtained as $-1.35\pm 0.29$ and $-1.29\pm 0.23$, respectively.
They are consistent with those previously measured in the frequency range above 10$^{-3}$ Hz.
\item 
In the 4--10 keV, and 10--20 keV bands, the fractional RMS variation $\eta$ observed in the HSS was
$\sim 2$ and $\sim 3$ times higher, respectively, than those measured in the LHS in the corresponding energies.
In the 2--4 keV band, $\eta$ in the HSS was comparable to that in the LHS.
\item 
In the LHS, $\eta$ weakly decreased towards higher energies.
This property is consistent with that seen in the short-term variability during the LHS.
\item 
In the HSS, $\eta$ was found to increase significantly with energy,
as a combination of  the following two effects.
One is that the disk component is stable while the hard tail varies.
The other is that the hard tail slope flattens as it becomes brighter.

\end{enumerate}

We have studied long-term variability of the hard tail component in the HSS.
This will provide important clues to the yet unidentified origin of this emission component.


This work was supported by RIKEN Junior Research Associate Program.
This work was  also partially supported by 
the Ministry of Education, Culture, Sports, Science and Technology (MEXT), Grant-in-Aid for Science Research 24340041,
and the MEXT Supported Program for the Strategic Research Foundation at Private Universities, 2014-2018.


\appendix
\chapter{PSD simulation of MAXI/GSC observation}
\label{gap}


{\bf A--1. How to simulate a MAXI/GSC light curve}

To estimate effects of the data gaps and the sampling windows in our Cyg X-1 observation with the MAXI/GSC,  
we make a simulated light curve in units of (counts s$^{-1}$ cm$^{-2}$), 
whose PSD has a form of $Af^{-1}$.
The normalization $A$ is chosen to be close to what has actually been observed with the MAXI/GSC.
Let us assume first that the observation is continuous with each data bin having a length $\Delta t$, 
and set it equal to a typical source transit time, $t_{\rm scan} = 54$ s.
According to the method described in \citet{Isobe} and \citet{Timmer}, the count rate $a_i$ at the $i$-th bin is generated as
\begin{eqnarray}
a_{i} = \frac{2}{\sqrt{N\Delta t}}\sum^{N_{F}-1}_{j=0}\{c_{i}(f_{j}) + s_{i}(f_{j})\} & \rm{(counts~~s^{-1}~~cm^{-2})} \nonumber \\
                                                       & \label{eq:A} \\
c_{i}(f_{j}) = R_{{\rm cos},j}\cos(2\pi f_{j}t_{i})   & \nonumber \\
s_{i}(f_{j}) = R_{{\rm sin},j}\sin(2\pi f_{j}t_{i})   & \nonumber \\
t_{i} = i\Delta t ~~~ (i = 0, ..., N-1)              & \nonumber \\
f_{j} = j\Delta f ~~~ (j = 0, ..., N_{\rm F}-1)      & \nonumber 
\end{eqnarray}
Here $N$ is the total time-bin number, 
of the order of $5\times 10^{5}$ (300d / 54s), 
$N_{\rm F} = N/2$ is the number of PSD bins, 
$\Delta f$ is the width of each PSD bin [$\Delta f = 1/(N\Delta t)$], $R_{{\rm cos},j}$ and $R_{{\rm sin},j}$ 
are independent gaussian-distributed random numbers with their mean values 0 and their standard deviations $\sqrt{Af_{i}^{-1}}$, 
while $c_i(f_j)$ and $s_i(f_j)$ are the corresponding Fourier components.
The DC component, $c_0$, is chosen so as to reproduce the average source count rate.
The factor $\frac{2}{\sqrt{N\Delta t}}$ is the coefficient arising from the inverse Fourier transformation.
Thus, \{$a_i;\ i = 1,2,...N$\} provides a simulated light curve, but without including the Poisson error.

\begin{figure*}
\begin{center}
   \includegraphics[width=16cm]{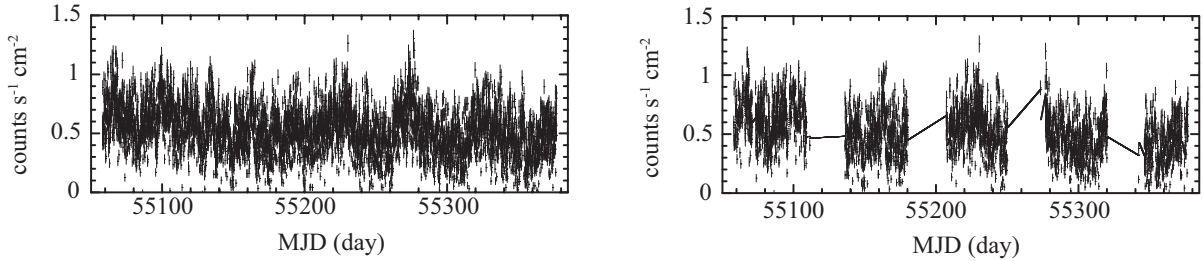}
 \end{center}
 \caption{The left figure is a simulated light curve calculated with equation (\ref{eq:Yi}).
 The right one is after applying the data gaps and filling them by linear interpolations.
 The sampling time is $\Delta t =$ 5400 s.}
\label{fg:simlc}
\end{figure*}

To simulate the Poisson noise, 
a Poisson distributed random number $b_i$ is generated at the $i$-th bin as
\begin{eqnarray} 
  b_i = p(a_{i}\cdot(S\cdot t_{\rm scan}) + \overline{B})~~~ \rm{(counts)}.
\label{eq:C}
\end{eqnarray}
Here $p(x)$ means a Poisson distributed random number with the mean value of $x$, 
$S$ (cm$^2$) is the average effective area per scan, 
$a_{i}\cdot(S\cdot t_{\rm scan})$ represents the simulated signal counts, 
and $\overline{B}$ denotes the average background counts per transit of a source.
We ignore scan-by-scan scatter in $S$, $t_{\rm scan}$, and $\overline{B}$.
From this $b_i$, we subtract $\overline{B}$, and divide the result by $S\cdot t_{\rm scan}$, 
to obtain a simulated data point $Y_i$ as
\begin{equation} 
 Y_i = (b_i - \overline{B})/(S\cdot t_{\rm scan})~~~ \rm{(counts~~s^{-1}~~cm^{-2})}.
  \label{eq:Yi}
\end{equation}
An example of the light curve obtained in this way is shown in figure \ref{fg:simlc}.
\\

\begin{figure*}
\begin{center}
   \includegraphics[width=12cm]{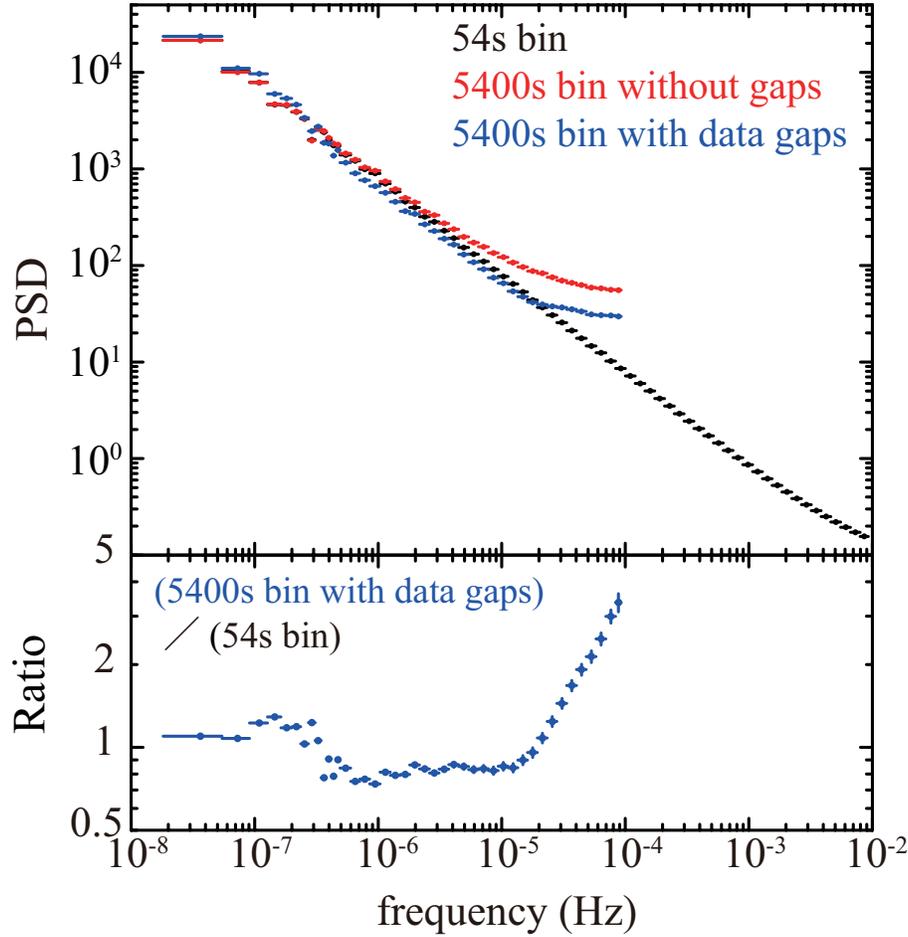}
 \end{center}
 \caption{Simulations of the alias effects and sampling effects on the PSD.
 The black data are the PSD calculated from 100 fake light curves, 
 each continuous with a bin width of 54s.
 The red one is the PSD calculated after sparse sampling 
 (once every 5400s), to the original light curves, and blue one further considering the data gaps that are present in the read data.
 The bottom panel shows the transfer-function, 
 obtained by the dividing the blue PSD by the black one.}
\label{fg:alias+gap}
\end{figure*}

We simulate one hundred light curves composed of continuous 54-s bins, with a total real observation time of 318 d  
which is the longest observation segment in the LHS (data No.1-all in table \ref{tb:hardem}).
These 100 light curves with 54-s bin were Fourier transformed into PSDs, 
which were averaged to give the PSD shown in black in figure \ref{fg:alias+gap}.
It recovers the assumed $\propto f^{-1}$ shape.

{\bf A--2. Sampling effect}

The MAXI exposure for a celestial object is far from being continuous.
A MAXI light curve consists of discrete snap-shot scans, 
which are separated by $\Delta t = 92 {\rm min}\simeq 5400 {\rm s}$ and lasting for $\sim t_{\rm scan}$ each.
In such sparse-sampling observations with $t_{\rm scan} \ll \Delta t$, significant alias appears in the high frequency end (subsection 3.2.2; \cite{Kirchner}).

To estimate the effects of aliasing (sampling effects), 
we retained one every 100th bin, 
and discarded the remaining 99 data points.
Analysis of these sparse light curves yielded the red PSD in figure \ref{fg:alias+gap}.
The red points have more power than the black one towards the Nyquist frequency.
This is because the power from $10^{-4}$ to $10^{-2}$ Hz, 
which was smeared out in the black PSD by averaging over the original 100 data points, 
now appears in frequencies below $10^{-4}$ Hz.
This excess represents the alias power.

{\bf A--3. Gap effect}

Next, to simulate the observational gaps, 
the sparse (5400s bin) light curves were multiplied with exactly the same sampling windows as in the actual data (data No.1-all).
An example is given in figure \ref{fg:simlc} (right).
After interpolating these gaps in the same manner as for the actual data, 
the fake light curves were again Fourier transformed.
The obtained PSDs were averaged, and are shown in figure \ref{fg:alias+gap} in blue.
The alias effect is still present.

{\bf A--4. Correction by a Transfer function}

The bottom panel of figure \ref{fg:alias+gap} shows the ratio of the PSD calculated from the sparse and gap-applied light curves, 
mimicking the actual observation, to the original PSD.
This is to be called the transfer function.
We corrected the sampling and gap effects in the MAXI/GSC observation, 
by dividing the PSD calculated from the real data by this transfer function.

To evaluate systematic errors associated with the transfer function, 
we also performed similar simulations by changing the PSD index to $-2$ and $-0.5$.
As expected, the transfer function became flatter ($\sim 5$ at 10$^{-5}$) for the PSD index $-2$ ($\sim 1$ at 10$^{-5}$ Hz), 
while more deviated from unity ($\sim 5$ at $10^{-5}$ Hz) when the PSD index is set at $-0.5$.
However, these effects are limited to frequencies above $\sim 10^{-5}$ Hz.


\chapter{Estimate of the Poisson noise}
\label{poisson}

The observed data are subject to statistical fluctuations, or Poisson noise.
Since this component is independent of the intrinsic source variability,
the derived raw PSDs are considered to be a direct sum of
the intrinsic and Poissionian contributions.
Although the Poisson noise would usually be a constant,
this is not true in the present case,
because of the data gaps, background variations, and other practical effects.
We hence estimate the Poisson-noise contributions
to the PSDs by a numerical simulation;
the results were already used in subtraction in section \ref{sec:psd}
when calculating the PSDs of figure \ref{fig:cygnpsd} and figure \ref{fig:avepsd}.

The aim of the above simulation is to construct light curves
that would be observed if Cyg X-1 had a constant intensity throughout.
This has been carried out in the following steps.
\begin{enumerate}
\item[1]
We start from calculating the mean intensity $\bar{y}$ of Cyg X-1
in a specified energy band,
as an average over the 5 years of the MAXI data.
\item[2]
The signal counts to be observed at the $i$-th scan
of ``non-varying Cyg X-1"
can be expressed as $\bar{y} (S \cdot  t_{\rm scan})_i$,
where the effective area $S$ and the source transit time $t_{\rm scan}$
are the same as in Appendix \ref{gap}.
\item[3]
The total counts to be detected in the $i$-th scan can be expressed as
$\bar{y} (S \cdot t_{\rm scan})_i + \alpha_i B_i$,
where $B_i$ is the background which was estimated scan-by-scan
and was already utilized in section \ref{sec:obs} to derive the light curves in figure \ref{fig:cygLC},
while $\alpha_i$ is a correction factor for the effective area in the $i$-th scan.
\item[4]
The above quantity is randomized into
$b_i \equiv p\left( \bar{y} (S \cdot t_{\rm scan} \right)_i + \alpha_i B_i)$
in the same way as in Appendix \ref{gap}.
Then, $\{b_i \}$ represent the simulated raw counts data,
before the background subtraction,
to be detected in this fake observation.
\item[5]
From this randomized counts $b_i$,
we subtract $\alpha_i B_i$  in the same way as in section \ref{sec:obs},
and divide the result by $(S \cdot t_{\rm scan})_i$,
to obtain the simulated light curve $\{Y_i \}$
which should be observed from ``non-varying Cyg X-1".
The procedure is summarized as
\begin{equation}
       Y_i = \left( b_i - \alpha_i B_i \right)/ \left( S \cdot t_{\rm scan} \right)_i
       ~~~ (\rm{counts~s}^{-1}~\rm{cm}^{-2})~.
\end{equation}
\end{enumerate}

\begin{figure}
 \begin{center}
  \includegraphics[width=8cm]{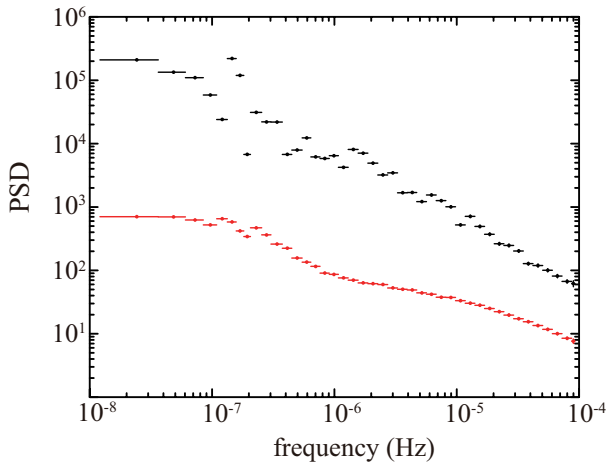}
	\hspace{30pt}
 \end{center}
 \caption{
The simulated 4--10 keV Poisson-noise contribution (red),
compared with the 4--10 keV PSD actually observed from Cyg X-1 in the HSS (black).
}
\label{fig:psd+pl}
\end{figure}

We produced the simulated light curves $\{ Y_i \}$ 100 times,
and calculated their PSDs in the same manner as for the actual data.
In comparison with the actual 4--10 keV PSD of Cyg X-1 in the HSS (black),
the 4--10 keV PSD thus simulated and averaged over the 100 runs
is shown in figure \ref{fig:psd+pl} in red.
Since the source intensity has been assumed to be constant,
the red PSD is considered to represent pure Poisson fluctuations.
While this PSD is approximately flat
on long time scales ($< 10^{-7}$ Hz)
as expected for white noise,
it decreases towards higher frequencies.
This is because the data gaps were filled with linear lines
which do not contain statistical fluctuations:
the Poisson noises  are diluted on time scales shorter than those of the data gaps.
In any event, the estimated Poisson-noise contribution is
$<15\%$ of the signal PSD at all frequencies analyzed here,
and particularly below a few percent at $<10^{-6}$ Hz.


\chapter{Mass accretion rates through optically thick accretion disk and corona}
\label{mdot}

\begin{figure*}[bhtp]
  \begin{center}
   \includegraphics[width=18cm]{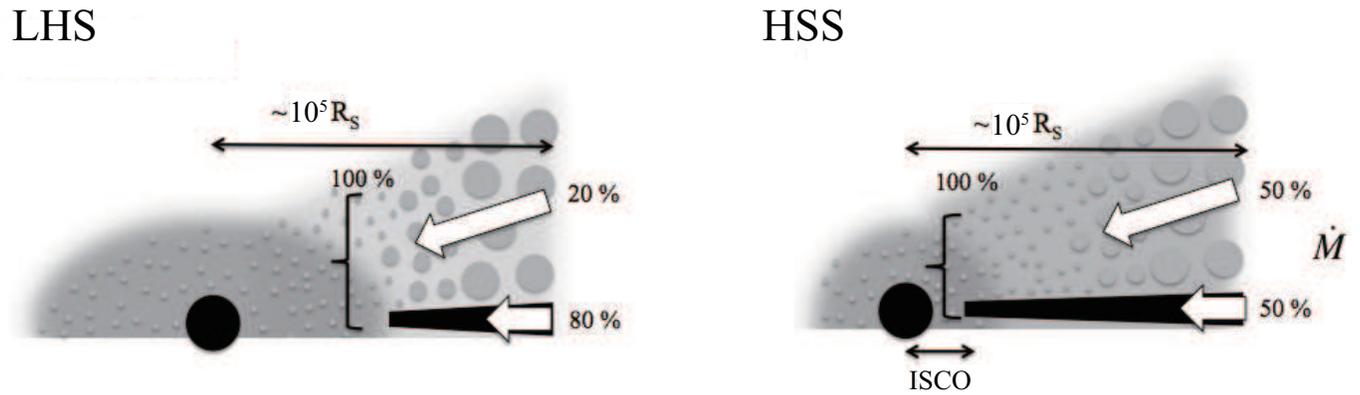}
  \end{center}
  \caption{Ilustrations of accretion models in the LHS (left) and in the HSS (right).
  The optically-thick accretion disk is shown with a thick black line, which is truncated at a certain radius in the LHS, 
  but extends down to the ISCO in the HSS.
  In both states, the optically-thin flow extends up to $\sim$ 10$^{5}R_{\rm s}$, 
  while it turns into the Comptonizing corona at small radii. 
  The percentages mean the fractional mass accretion rate in each state.}
\label{fig:cartoon}
\end{figure*}

Utilizing schematic drawings in figure \ref{fig:cartoon}, 
we consider an accretion model of Cyg X-1 that can explain our results.
The optically-thick accretion disk (thick black line) is assumed to extend from outer regions inwards, but is truncated at a certain radius in the LHS.
Meanwhile, in the HSS it is thought to extend down to the ISCO.
In both states, the optically-thin flow is assumed to extend from an outer region ($\sim 6 \times 10^{5} R_{\rm s}$) to the ISCO.
The radii, observed low frequency variations of the mass flow are assumed to be 
generated in the optically-thin flow at large radii, 
and propagates through the optically-thin flow to the vicinity of the BH to make the hot corona vary on these long time scales.

After \citet{chura}, let us consider mass accretion rate through the optically-thick disk; $\dot{M}_{d}$ 
and through the optically-thin flow; $\dot{M}_{c}$.
The total accretion rate is given as $\dot{M}_t = \dot{M}_d + \dot{M}_c$.
We further assume that 
the Comptonized component in the energy spectrum is mainly powered by $\dot{M}_{c}$, 
while the disk emission by $\dot{M}_{d}$.
Let us describe fractional variations in $\dot{M}_{c}$ and $\dot{M}_{d}$ as $f_c$ and $f_d$, respectively, 
and assume that these quantities depend neither  on the frequency nor on the spectral state.
Then, the observed normalized power should be proportional to $[(f_{c}\dot{M}_{c})^2+(f_{d}\dot{M}_{d})^2]/(\dot{M}_{t})^2$.
When the optically-thick flow is not variable, $f_{d}$ is equaled to zero.

In the LHS, the optically-thick accretion disk is truncated at a certain radius, 
and within that radius, only the optically-thin flow exists ($\dot{M}_{t} = \dot{M}_{c}$; \cite{chura}). 
Then, since the time variation above 0.1~Hz originates only from the optically-thin flow,  
the normalized power of this range is $f_{c}^2$ itself; NPSD $= f_{c}^2$.
On the other hand, in the frequency less than 10$^{-3}$~Hz the NPSD is roughly $\frac{1}{25}$  of the extrapolation of the NPSD above 0.1~Hz.
Therefore $\bigl[\frac{(f_{c}\dot{M}_{c})^2}{(\dot{M}_{t})^2}\bigr]_{\rm LHS} = \frac{1}{25} f_{c}^{2}$, 
where $f_{d}=0$ is assumed.
The suffix LHS specifies the states.
Since we assume that $f_{c}$ is independent of frequency, 
we obtain  $\bigl[\frac{\dot{M}_{c}}{\dot{M}_{t}}\bigr]_{\rm LHS} \sim\frac{1}{5}$, 
and hence $\bigl[\frac{\dot{M}_{d}}{\dot{M}_{t}}\bigr]_{\rm LHS} \sim\frac{4}{5}$.
In the hypothetical model, 
the PSD flattening at $10^{-3} - 10^{-1}$ Hz (figure \ref{fig:hkkchura}) is explained that 
the higher-frequency variations ($> 10^{-1}$ Hz) are produced inside the disk-truncation radius, 
while the lower-frequency ones ($< 10^{-3}$ Hz) mainly originate in the region where the two flows co-exist.

Similar arguments can be performed on the HSS.
In the low frequency region below $10^{-4}$~Hz, the normalized power of the HSS, 
$\bigl[ \frac{(f_{c}\dot{M}_{c})^2}{(\dot{M}_{t})^2} \bigr]_{\rm HSS}$, 
is about six times larger than 
that in the LHS, 
$\bigl[ \frac{(f_{c}\dot{M}_{c})^2}{(\dot{M}_{t})^2} \bigr]_{\rm LHS} = \frac{1}{25} f_{c}^{2}$ (see figure \ref{fig:hkkchura}).
Since we assume that $f_{c}$ does not depend on the states, 
$\bigl[\frac{(f_{c}\dot{M}_{c})^2}{(\dot{M}_{t})^2}\bigr]_{\rm HSS} = \frac{6}{25} f_{c}^{2}$. 
Then, $\bigl[\frac{\dot{M}_{c}}{\dot{M}_{t}}\bigr]_{\rm HSS} = \sqrt{\frac{6}{25}} \sim \frac{1}{2}$, 
and hence $\bigl[\frac{\dot{M}_{d}}{\dot{M}_{t}}\bigr]_{\rm HSS} \sim\frac{1}{2}$.
These values are supported by the fact that  
the luminosity of the disk component is roughly the same as that of the Comptonized component in the HSS (see table \ref{tb:especcomp}).
The inferred fractions of the mass accretion rate through the assumed two flows are summarized in figure \ref{fig:cartoon}.

\end{document}